# Room temperature entanglement between distant single spins in diamond.

F.Dolde[1], I.Jakobi[1], B.Naydenov[1,2], N.Zhao[1], S.Pezzagna[3], C.Trautmann[4,5], J.Meijer[3], P.Neumann[1], F.Jelezko[1,2], and J.Wrachtrup[1]

[1] 3. Physikalisches Institut, Research Center SCoPE, and IQST, Universität Stuttgart, Germany

[2] Institut für Quantenoptik, and IQST, Universität Ulm, Germany

[3] RUBION, Ruhr Universität Bochum, 44780 Bochum, Germany

[4] GSI Helmholtzzentrum für Schwerionenforschung, 64291 Darmstadt, Germany

[5] Technische Universität Darmstadt, Darmstadt, Germany

**Entanglement is the central yet fleeting phenomena of quantum physics. Once being considered a peculiar counter-intuitive property of quantum theory it has developed into the most central element of quantum technology[1] providing speed up to quantum computers, a path towards long distance quantum cryptography and increased sensitivity in quantum metrology. Consequently, there have been a number of experimental demonstration of entanglement between photons[2], atoms[3], ions[4] as well as solid state systems like spins or quantum dots[5-7], superconducting circuits[8,9] and macroscopic diamond[10]. Here we experimentally demonstrate entanglement between two engineered single solid state spin quantum bits (qubits) at ambient conditions. Photon emission of defect pairs reveals ground state spin correlation. Entanglement (fidelity = 0.67 ± 0.04) is proven by quantum state tomography. Moreover, the lifetime of electron spin entanglement is extended to ms by entanglement swapping to nuclear spins, demonstrating nuclear spin entanglement over a length scale of 25 nm. The experiments mark an important step towards a scalable room temperature quantum device being of potential use in quantum information processing as well as metrology.**

Engineering entangled quantum states is a decisive step in quantum technology. While entanglement among weakly interacting systems like photons has been demonstrated already in the early stages of quantum optics, deterministic generation of entanglement in more complex systems like atoms or ions, not to speak of solids, is a relatively recent achievement[11]. Usually in solid state systems rapid dephasing ceases any useful degree of quantum correlations. Either decoupling must be used to protect quantum states or careful materials engineering is required to prolong coherence. Most often however, and this is especially important for solid state systems, one needs to resort to low (mK) temperatures to achieve sufficiently robust and long lasting quantum coherence. Only spins are sufficiently weakly coupled to their environment to allow for the observation of coherence at room temperature[12].

Diamond defect spins are particularly interesting solid state spin qubit systems. A number of hallmark demonstrations like single, two and three qubit operations, high fidelity single shot readout[13], one and two qubit algorithms[14,15] as well as entanglement between nuclear and electron and nuclear spin qubits have been achieved[6,16]. Different schemes to scale the system to a larger number of entangled electron spins have been proposed[17-19]. A path towards room temperature entanglement is strong coupling among the ground state spin magnetic dipole moment of adjacent defects centers. This mutual dipolar interaction scales as distance $d^{-3}$ (Fig. 1d) and should be larger than the interaction of each electron spin to the residual paramagnetic impurities or nuclear spin moments in the lattice. Typical cut-off distances for strong interaction are thus limited by the electron spin dephasing time (ms) to be around 30 nm. Here we demonstrate entanglement between two electron and nuclear spins at a distance of approximately 25 nm. At these distances magnetic dipole coupling is strong enough to attain high fidelity entanglement while being able to address the spins individually by super-resolution optical microscopy[20].

The optical as well as spin physics of nitrogen vacancy (NV) defects in diamond has been subject to numerous investigations[11,21]. The fluorescence intensity of the strongly allowed optical transition between ground and excited spin triplet state depends on the magnetic quantum number of the ground state and it is larger for $m_S = 0$ and smaller for $m_S = \pm 1$, allowing optical read-out of the electron spin[22]. The coherence time $T_2$ of the NV$^-$ electron spin depends on the concentration of $^{13}$C spins and reaches up to 3 ms for $^{12}$C enriched diamond[11].

NV defect are either formed by nitrogen incorporation during growth or by implantation of nitrogen into high purity diamond material with a subsequent annealing step[23]. Here we have chosen the latter method to generate proximal diamond defect pairs. To generate strongly coupled defect pairs with high probability and at the same time optimum decoherence properties nitrogen ions ($^{15}$N$^+$) with kinetic energies of 1 MeV, corresponding to an implantation depth of 730 nm have been implanted using a 10 μm thick mica nano-aperture mask (hole diameter 20 nm). This process creates NV pairs at distances less than 20 nm with a success rate of 2 % (see supplementary information)[24].

Fig. 1b shows the spin energy levels of the NV pair together with the electron spin resonance (ODMR) spectrum of two coupled electron spins. The spectrum in secular approximation is described by

$$H = \sum_{i=NV\,A}^{NV\,B}(\hat{S}_i \underline{\underline{D_i}} \hat{S}_i + \gamma_e \underline{B} \hat{S}_i) + \nu_{dip} \hat{S}_{zA} \hat{S}_{zB}, \quad (1)$$

where $D$ is the zero field splitting, $B$ the magnetic field, $\gamma_e$ the gyromagnetic ratio and $S_{A(B)}$ the spin operator of NV A(B). Electron spin flipflop terms like ($S_{xA}S_{xB}$, $S_{yA}S_{yB}$) can be neglected as long as energy splittings are dominated by axial fine structure interaction larger than $\nu_{dip}$ (see supplementary information). The two defects are oriented along two different directions of the diamond lattice and hence the orientation dependence of the Zeeman term allows for individual addressing by different microwave frequencies. To investigate the magnetic dipolar coupling between the two defects we induce spin transitions ($\Delta m_S = \pm 1$) on both defects and use NV A as a sensitive magnetometer[25] to measure spin flip induced changes in the magnetic dipole field of NV B yielding a dipolar coupling constant of $\nu_{dip} = 4.93 \pm 0.05$ kHz (Fig. 1e). This coupling strength can be significantly enlarged by using the qutrit nature of the two spins in the pair and inducing $\Delta m_S = \pm 2$ (double quantum transitions, DQ) on both NVs yielding $\nu_{dip} = 19.72 \pm 0.2$ kHz. The measured values for dipolar interaction allow a maximum distance of $29.6 \pm 1.4$ nm between the two defects. The actual distance obtained by involving microwave assisted super-resolution microscopy yields

$25 \pm 2$ nm. We would like to note that the coupling did not change over month indicating the room temperature stability of the defect pair.

To create high fidelity entanglement, strong coupling has to apply i.e. $\nu_{dip} > 1/T$ (where $T$ is the relevant coherence time). The present moderate coupling, is masked by spectral diffusion of the two individual electron spins ($T_{2\ NV\ A}^* = 27.8 \pm 0.6$ μs and $T_{2\ NV\ B}^* = 22.6 \pm 2.3$ μs) i.e. $\nu_{dip} < 1/T_2^*$. This limitation can be overcome by eliminating low frequency environmental noise components through additional refocusing steps in the entanglement process resulting in a new lower limit for strong coupling $\nu_{dip} > 1/T_2$. The electron spin relaxation and coherence times of the two NVs are $T_1 = 1.12 \pm 0.26$ ms, $T_{2\ NV\ A\ DQ} = 150 \pm 18$ μs and $T_{2\ NV\ B\ DQ} = 514 \pm 50$ μs. $T_{2\ DQ}$ is sufficiently long to allow for creation of entanglement as described below.

After optically initializing the system in $|m_{S\ A}, m_{S\ B}\rangle = |00\rangle$ a π/2 rotation on both NVs leads to $\frac{1}{2}(|00\rangle - |10\rangle - |01\rangle + |11\rangle)$. Under the influence of mutual dipolar coupling the system is evolving freely for a time τ resulting in a state dependent phase acquisition $\frac{1}{2}(|00\rangle - |10\rangle - |01\rangle + e^{-i\varphi}|11\rangle)$ where $\varphi = 2\pi\nu_{dip}\tau$ is the correlated phase due to dipolar interaction. After a π rotation and an additional free evolution period τ a second phase is accumulated $\frac{1}{2}(e^{-i\varphi}|00\rangle + |10\rangle + |01\rangle + e^{-i\varphi}|11\rangle)$. With a final π/2 rotation the accumulated phase is mapped onto $\frac{1}{2}((e^{-i\varphi} - 1)|00\rangle + (e^{-i\varphi} + 1)|11\rangle)$. For $\tau = \frac{1}{4\ \nu_{dip}}$ this is $\Phi^+ = \frac{1}{\sqrt{2}}(|00\rangle + i|11\rangle)$, a maximally entangled Bell state (for details see supplementary information).

Fig. 2b shows the state evolution upon application of the entanglement gate as a function of interaction time τ. The blue line is a simulation of the entanglement gate using Hamiltonian (1) with coherence times taken from experiments. For $\tau = 12.5$ μs the state has evolved to $\Phi^+$. Using local operations this state can be transformed into a set of different entangled states e.g. two π pulses transform $\Phi^+$ to $\Phi_{DQ}^+ = \frac{1}{\sqrt{2}}(|-1-1\rangle + i|11\rangle)$. To exemplify the entangled nature of the two defect spins further a full spin state tomography was performed for the $\Phi_{DQ}^+$ state. In Fig. 2c and d the reconstructed density matrix is shown yielding a state fidelity of $0.67 \pm 0.04$ (for a detailed description see supplementary information).

Entanglement between spins is also inferred from fluorescence emission properties of the entangled defect pair. The steady state fluorescence emission of $\Psi^+ = \frac{1}{\sqrt{2}}(|01\rangle + i|10\rangle)$, $\Phi^+ = \frac{1}{\sqrt{2}}(|00\rangle + i|11\rangle)$ as well as a correspondingly separable spin state of both NV centers (e.g. $\frac{1}{2}(|00\rangle - |10\rangle - |01\rangle + |11\rangle)$) is identical. However two photon correlations reveal a difference between spin entangled and mixed states. A Φ state will lead to a higher probability of two photon emission than an uncorrelated superposition state whereas the Ψ state results in a lower probability, respectively. In Fig. 3 two photon correlation measurements and the corresponding classical correlations are shown.

The lifetime of the entangled states is limited by electron spin dephasing measured to be $T_{2\ NV\ A}^* = 27.8 \pm 0.6$ μs and $T_{2\ NV\ B}^* = 22.6 \pm 2.3$ μs. The measured entanglement lifetime is $T(\Phi_{DQ}^-) = 28.2 \pm 2.2$ μs and $T(\Psi_{DQ}^-) = 23.7 \pm 1.7$ μs (Fig. 4c and supporting online information). It is interesting to note that the lifetimes for states $\Phi^\pm$ and $\Psi^\pm$ are identical although $\Psi^\pm$ is known to constitute a decoherence free subspace for magnetic field noise dominated dephasing processes[26].

However, cancellation of decoherence effects in $\Psi^\pm$ only occurs if the magnetic field noise is identical for both NV A and NV B. Apparently this is not the case for the pair. One reason is the different orientation of the pair NVs with respect to $B_0$ which would result in non-ideal decoherence free subspaces. In addition from a previous analysis of spin dephasing in diamond defect centers it became clear that electron spin dephasing is dominated by nuclear spins in the vicinity of the defect. From those studies it was concluded that nuclei with distance on the order of a few nm dominate decoherence of the electron spin. Since the two defects are separated by about 25 nm it is evident that each NV is dephased by a separate set of nuclear spins with no correlation being present.

In order to store entanglement for a longer period, we designed an experimental scheme (Fig. 4a) to transfer electron spin entanglement to $^{15}$N nuclear spins of the NV. Instead of swapping entanglement by driving nuclear spins directly[6] or using excited state spin dynamics[27], we used a combination of a nonaligned static magnetic field and selective gates[28] on the electron spins to generate electron nuclear and at the end of the pulse sequence purely nuclear spin entanglement. The swap operation of the population and coherences i.e. entanglement transfer has a theoretical efficiency of 1, i.e. all electron spin entanglement is converted into nuclear spin entanglement (see supplementary information). Limited pulse (i.e. gate) efficiency however results in an efficiency of around 41% for storage and retrieval in experiments. As shown in Fig. 4c the lifetime of the entangled nuclear spin state however is drastically longer than that of the electron spin state with an effective storage time of over one ms. A comparison with the electron spin relaxation time (shown in Fig. 4c demonstrates that it is the electron $T_1$ which limits the entanglement in the nuclear spin quantum register. We would like to stress that during this time there is no nuclear spin interaction, i.e. the two nuclear spins are entangled without interaction (a few mHz compared to a lifetime of ms). Therefore, this is deterministic remote entanglement of two solid state nuclear spins. Decoherence of the stored entangled state could be further suppressed by repolarizing the electron spin (i.e. lengthening the effective electron spin $T_1$) allowing nitrogen nuclear coherence times beyond $T_2^* = 7.25$ ms[29]. By continuous strong optical excitation of both NV centers further improvement of entanglement storage into the range of seconds seems feasible[30].

The experiments presented in this paper mark a first step towards scaling room temperature diamond quantum register by demonstrating deterministic entanglement of electron and nuclear spins over some 10 nm distance. With the advent of diamond defect center nanotechnology more efficient generation of defect pairs and larger defect arrays seems to be tractable. For example by decreasing the implantation energy to about 10 keV and using the present mask technology should allow for a pair creation efficiency of almost 100 %. Recently techniques like nano implantation with positioning accuracies of 20 nm [23,31] and shallow implanted defect showing dephasing times not degenerated by surface proximity [32-34] have improved considerably. With the aid of those techniques controlled generation of large scale arrays seem to be within reach, paving the way towards room temperature quantum devices.

## METHODS SUMMARY

The pair was produced by ion implantation in isotope enriched $^{12}$C diamond (99.99 %) using a specially designed mica mask with high aspect ratio apertures (1:100)[24]. Ground state depletion imaging (GSD) was performed to identify suitable candidates, which were investigated with Double-Electron-Electron-Resonance (DEER). The sample was investigated in a home build confocal microscope. For coherent control microwave radiation was synthesized (Rhode-Schwarz SMIQ 03B) and modulated by an IQ mixer with an arbitrary waveform generator (Tektronix AWG 520). The microwave was applied via a microstructure forming a split ring resonator lithographically fabricated on the diamond surface.

## ACKNOWLEDGEMENTS

The authors would like to acknowledge financial support by the EU via SQUTEC and Diamant, as well as the DFG via the SFB/TR21, the research groups 1493 "Diamond quantum materials" and 1482 as well as the Volkswagen Foundation. We thank R. Kolesov, R. Stöhr, G. Waldherr, S. Steinert, T. Staudacher, J. Michl, C. Burk, E6, H. Fedder, F. Reinhard and F. Shi for discussions and support.

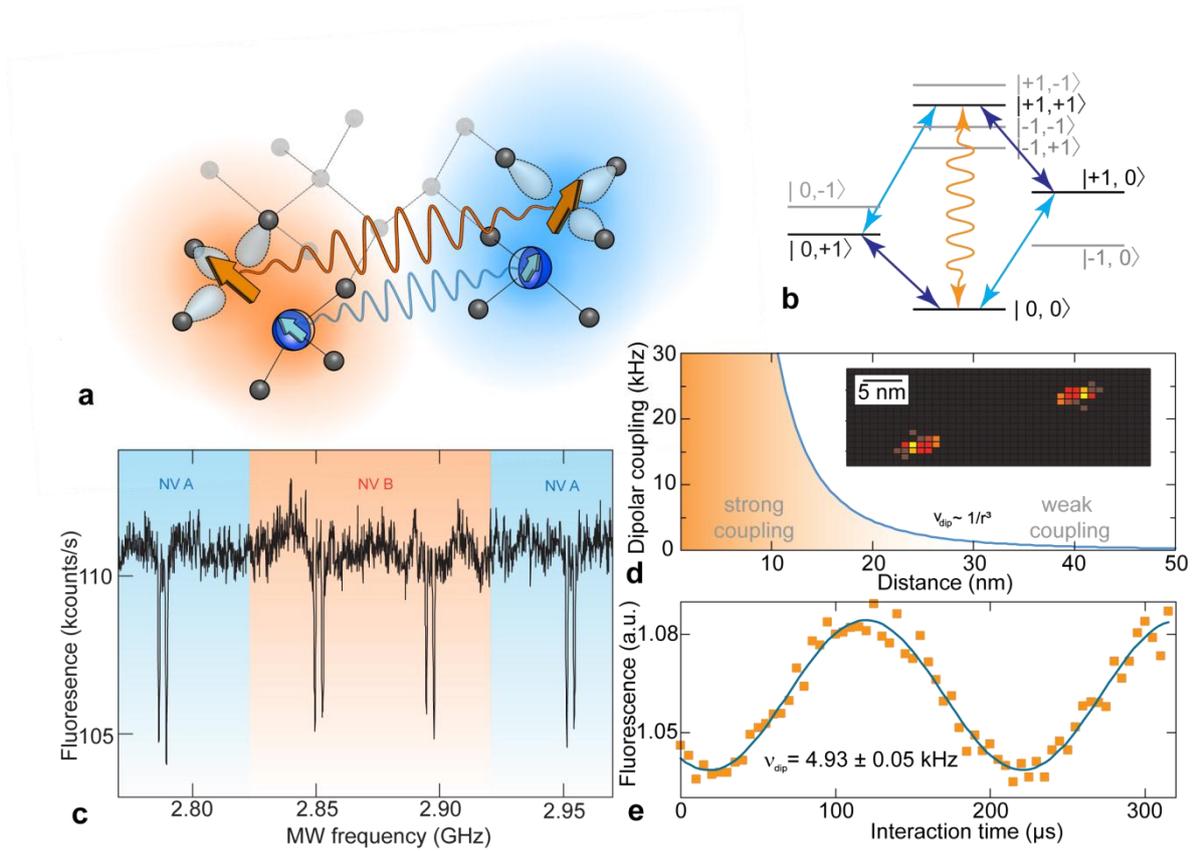

**Figure 1 | NV pair characteristics. a,** Schematics of the NV pair. The two NV centers have different orientations and a distance of d≅25±2 nm. The magnetic field is aligned with the axis of NV A. **b,** Level scheme of the combined system of two NV electron spins. Spin transitions with $\Delta m_s = \pm 1$ can be driven with microwaves (blue arrows). Transitions can be addressed individually because different magnetic field alignments result in different Zeeman shifts (see panel c). **c,** ODMR-spectrum of the NV pair at a magnetic field of $B = 32$ G. Two spin transitions can be attributed to each NV center. Both show a $^{15}$N (I = 1/2) hyperfine structure indicating that they stem from implanted nitrogen. **d,** The coupling strength $\nu_{dip}$ of two NV electron spins as a function of distance. With coherence times of $T_2 \cong 1$ ms strong coupling between defect spins can be achieved for $d < 30$ nm. Inset, optically resolved lateral distance between the two NV centers obtained by microwave assisted super-resolution microscopy. The measured lateral distance is $21.8 \pm 1.7$ nm (see supplementary information). **e,** Double-electron-electron-resonance experiment on the dipolar coupled NV pair. The oscillation shown is a direct measure of the coupling frequency $\nu_{dip} = 4.93 \pm 0.05$ kHz.

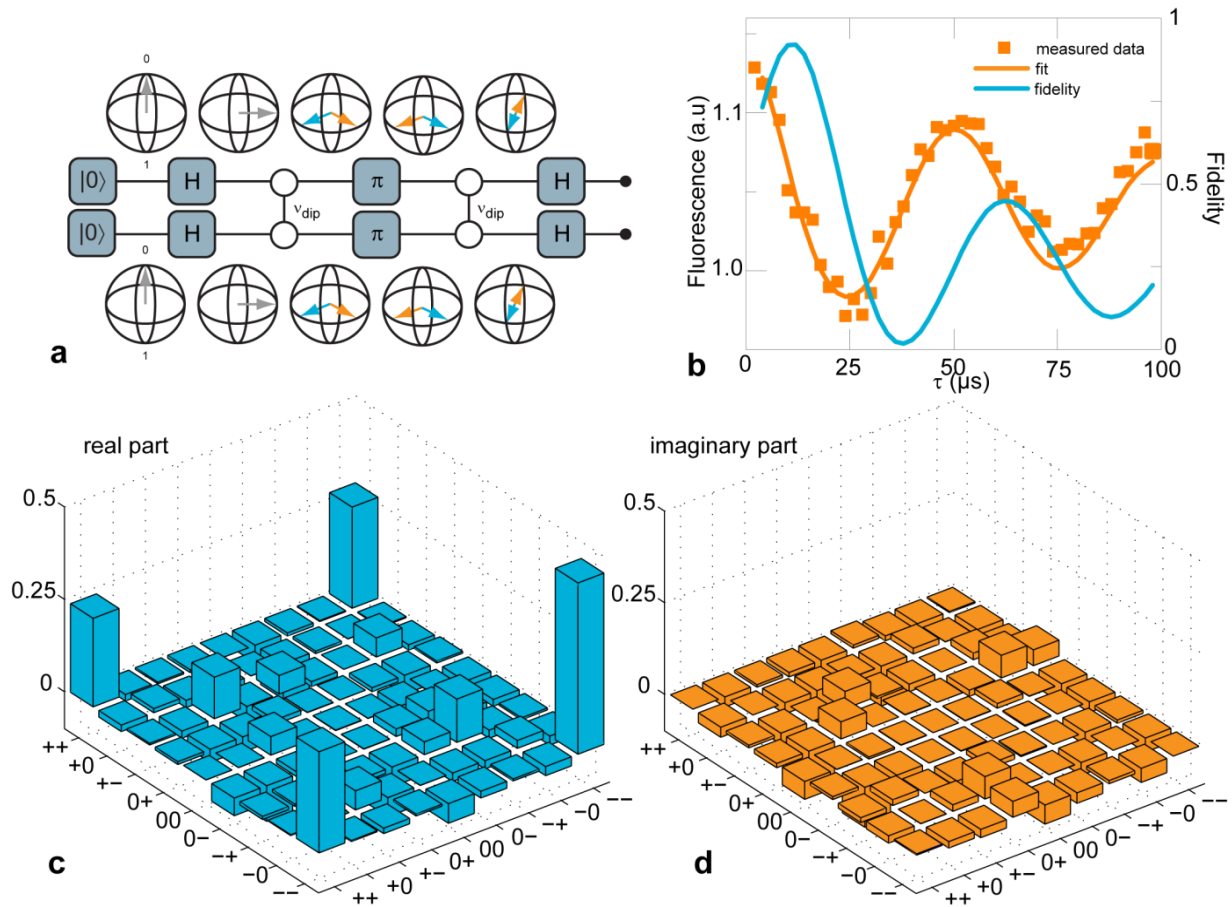

**Figure 2 | Bell state tomography**. **a**, Quantum circuit diagram of the entanglement scheme. A spin echo sequence on both individual spins reduces the effect of local noise while preserving the spin spin interaction. The latter realises a controlled phase gate which finally leads to the Bell state $|\Phi^-\rangle$ after an evolution time of $1/8\, \nu_{dip}$. **b**, Final state of entanglement scheme as a function of evolution times τ. The graph includes the simulated fidelity of reaching $|\Phi^-\rangle$. **c, d,** Real and imaginary part of the density matrix tomography of the $\Phi^+_{DQ} = \frac{1}{\sqrt{2}}(|-1-1\rangle + i|11\rangle)$ state.

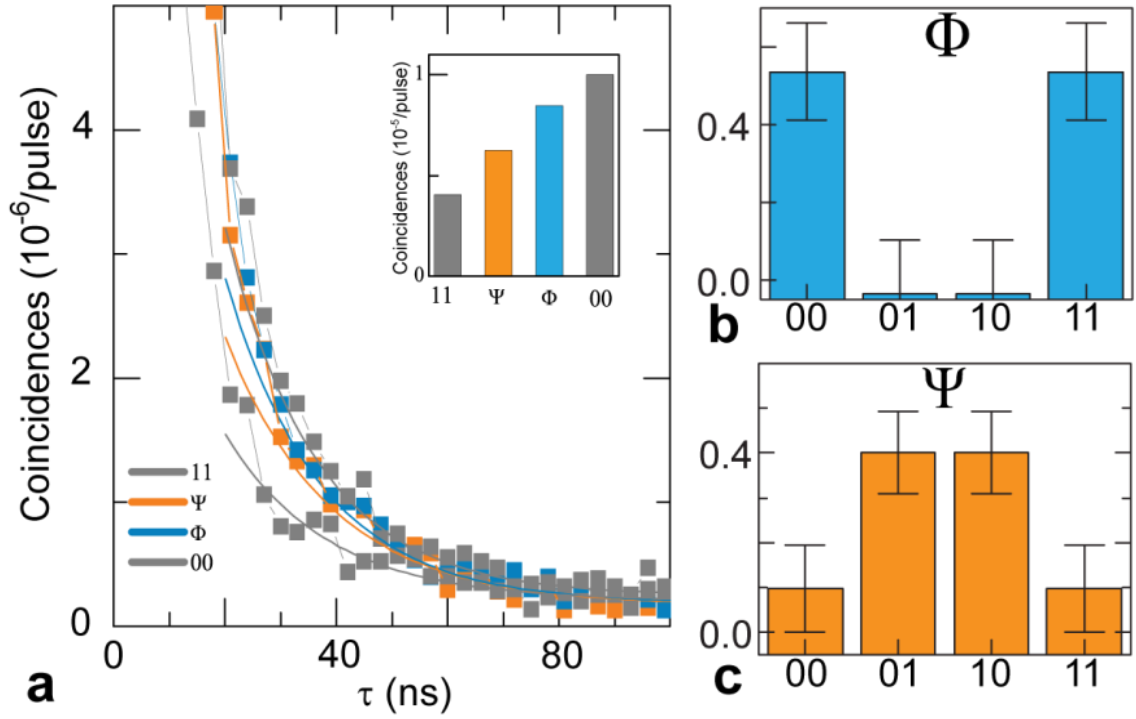

**Figure 3 | Two photon correlation measurements.** **a,** Results of two photon correlation measurement for entangled and mixed states. Photons close to the zero delay have been discarded. The inlay shows the fitted amplitude of two photon coincidences at τ = 20 ns. **b,** Reconstructed population correlation of a $\Phi^+ = \frac{1}{\sqrt{2}}(|00\rangle + i|11\rangle)$ state in a reduced basis of $m_S = 0$ and $m_S = +1$. The fidelity of the main diagonal is $F(\Phi^+_{class}) = 1.07 \pm 0.19$. **c,** For $\Psi^+ = \frac{1}{\sqrt{2}}(|01\rangle + i|10\rangle)$ the main diagonal fidelity is $F(\Psi^+_{class}) = 0.81 \pm 0.15$. (see supplementary information for details)

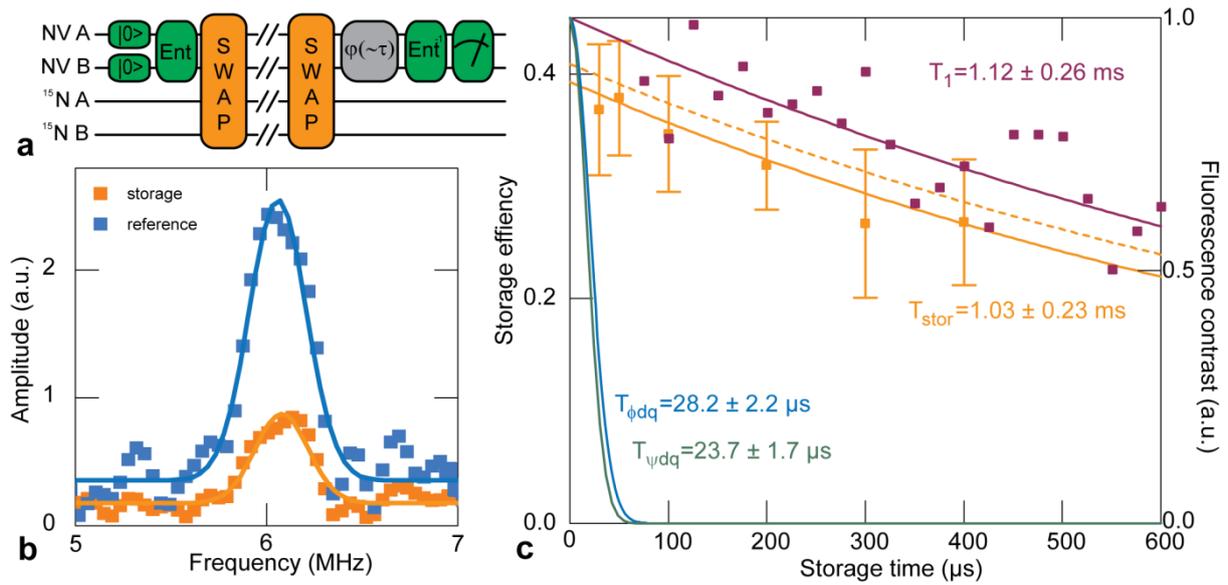

**Figure 4 | Entanglement storage in $^{15}$N. a,** Entanglement storage scheme. Selective π pulses creating a swap operation store the entanglement in the nitrogen nuclear spin. **b,** FFT of the entangled states' collective phase evolution after entanglement storage (orange) and a reference measurement without entanglement storage (blue). **c,** The dependence of the entanglement recovery efficiency on the storage time is shown in orange. The solid orange line is the exponential fit to the data whereas the orange dashed line is the simulated storage and retrieval efficiency given the imperfect SWAP gate. Magenta dots and line are the measurement and fit of $T_1$. The blue and green lines are entanglement lifetimes without storage.

# Supplementary information

# for

# "Room temperature entanglement between distant single spins in diamond."

F. Dolde[1], I. Jakobi[1], B. Naydenov[1,2], N. Zhao[1], S. Pezzagna[3], C. Trautmann[4], J. Meijer[3], P. Neumann[1], F. Jelezko[1,2], and J. Wrachtrup[1]

[1] 3. Physikalisches Institut, Research Center SCoPE, and IQST, Universität Stuttgart

[2] Institut für Quantenoptik, and IQST, Universität Ulm

[3] RUBION, Ruhr Universität Bochum, 44780 Bochum, Germany

[4] GSI Helmholtzzentrum für Schwerionenforschung, 64291 Darmstadt, Germany

# Table of Contents

# Ion implantation

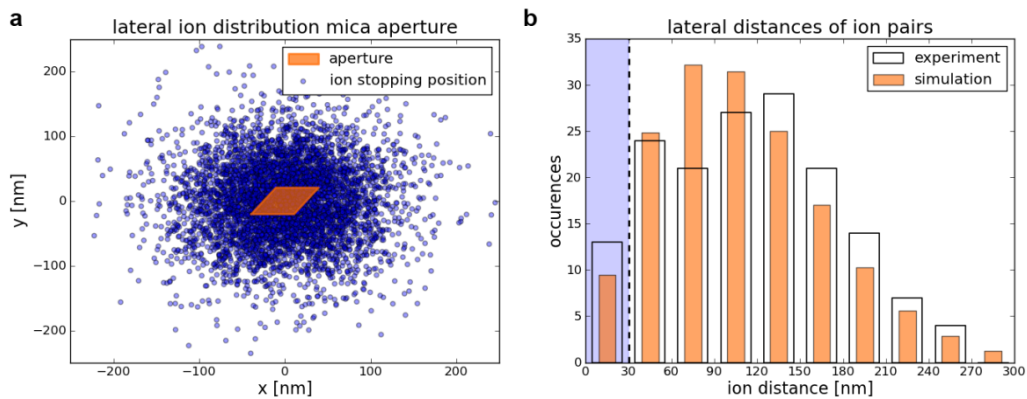

**Supplementary Figure 1 | SRIM simulation of 1 MeV $^{15}$N-ions in diamond. a.** lateral stopping positions (blue) of 5000 ions originating from a nano-channel in mica channel (orange). The flux through the 50 by 40 nm sized aperture is assumed to be homogeneous. The straggle is 118.9 nm. **b.** lateral distances of NV pairs were measured using super resolution microscopy (transparent). For comparison a histogram of distances of simulated ion pairs is shown (orange).

So far there are two major implantation methods. There is focused ion beam implantation allowing for MeV implantation energies generating deep implanted NV centres with long coherence times but only achieve a spatial resolution in the order of a few μm[1]. Implantation through a mask is capable of generating very implantation with precision on the order of a few tens of nanometers[2,3] but due to the thin mask material only allows for implantation energies in the keV range. In this work we used a high aspect ratio nano channel mask in mica (aspect ratio > 1:160). The mask was created by bombardment of the mica sheet with 1.7 GeV Samarium ions and etching of the ion track in HF resulting in 50 by 40 nm sized channels in a 8 μm thick mica slab[4]. This thick mica layer with the outstanding aspect ratio allows for MeV implantation energies without losing special resolution due to the beam waist.

The implantation of 1 MeV $^{15}$N ions was simulated using the "SRIM" software pack[5]. For that purpose the flux through the mica channel was assumed to be homogeneous. Implanted ions (Supplementary Fig. 1 a) have a distribution with a straggle of 118.9 nm. The ratio of ion pairs with distances below 30 nm is 1.97 %. Thus a reasonable number of implanted NV pairs are expected to show dipolar coupling while still having decent coherence times.

1 MeV $^{15}$N ions were implanted through mica channels using a Tandem accelerator. By implanting $^{15}$N with a natural abundance of 0.37 % it was possible to confirm that the NV⁻ centres found stem from implanted nitrogen ions[6]. Implanted samples were annealed at 800°C for 8 hours. To increase the intrinsic coherence times, a 99,9 % $^{12}$C sample was used allowing for coherence times in the order of ms[7]. Defect distances within NV⁻ pairs were measured using ground state depletion microscopy

(GSD)[8]. Supplementary Fig. 1 b reveals a good agreement of the lateral NV-NV distance histogram for simulation and the measured results.

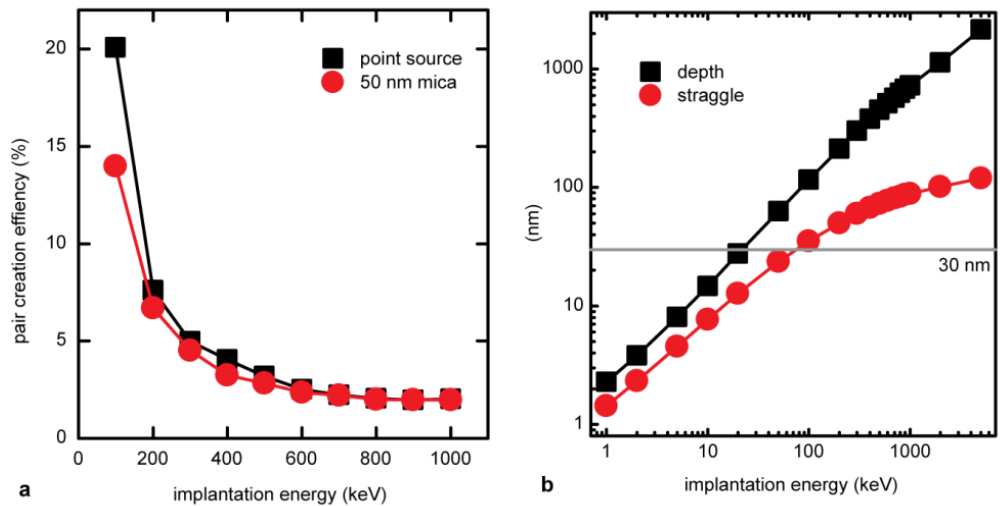

**Supplementary Figure 2 | NV pair creation efficiency. a.** percentage of NV pairs created for a point source (blue data) and using a 50 nm mica nano-aperture simulated using SRIM **b.** simulated depth and straggle for different implantation energies using SRIM

With a decrease of the implantation energy the pair creation efficiency can be increased even further. For depth of a few 100 nm coherence times in the order of a few hundred µs have been reported[9]. Here the approximation of the mica mask as a point source does not hold anymore (see Supplementary Fig. 2 a). However with the mica masks used for this experiments creation efficiencies of 14 % are possible and mask dimensions below 20 nm have been reported[4]. With further improvement of surface treatments allowing for long coherence times close to the surface[10-12] a deterministic creation using low implantation energies will be feasible.

# Charge State detection

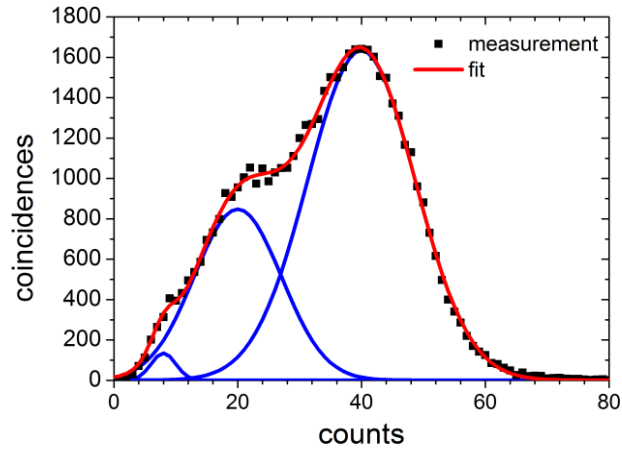

**Supplementary Figure 3 |** Histogram of detected counts under yellow (597.5nm) laser irradiation. The photon count time was 5 ms.

The NV centre exists in two different charge states. The ration of these two states was measured to be 30:70 $NV^0$:$NV^-$ [13] for NVs created during the crystal growth. Since the neutral charge state is a spin 1/2 system, the entanglement sequence is not successful if one or both NV centres are in this state. This gives rise to a single entity phase in time evolution measurements (the 3 MHz peak in Supplementary Fig. 8). Fortunately the zero phonon line of the neutral charge state (575 nm) and the negative charge state (638 nm) are separated by approximately 60 nm in wavelength, allowing us to address only the negative charge state with excitation light between 580 and 638 nm. Because the neutral NV centre cannot be excited in this wavelength range, all detected fluorescence photons in this excitation bandwidth can be associated with the negative charge state.

The ionisation of the NV centre is induced by a two photon absorption process. If the laser power is reduced significantly (about 1 ‰ of the saturation intensity) the cross section of two photon events is reduced sufficiently to allow for detection of fluorescence photons associated with the negative charge state only.

With the usage of a FPGA (Field Programmable Gate Array) counting device a threshold can be introduced to preselect only measurement sets where both NVs are negatively charged.

# Spin-spin coupling

### Model and Hamiltonian

The system of two NV centres is described by the Hamiltonian

$$H = H_A + H_B + H_{dip}, \qquad (1)$$

where $H_A$ and $H_B$ are the Hamiltonians of the two independent NV centres, NV A and NV B, respectively and $H_{dip}$ describes the dipolar interaction between them. The two NV centres have different orientations. Their Hamiltonians (neglecting strain) are

$$H_A = \Delta(S_z^A)^2 - \gamma_e \boldsymbol{B} \cdot \boldsymbol{S}_A + a_N \boldsymbol{S}_A \cdot \boldsymbol{I}_A, \qquad (2)$$

$$H_B = \Delta(S_{z'}^B)^2 - \gamma_e \boldsymbol{B} \cdot \boldsymbol{S}_B + a_N \boldsymbol{S}_B \cdot \boldsymbol{I}_B \,, \tag{3}$$

where $\hat{z} \parallel [111]$, $\hat{z}' \parallel [1\bar{1}\bar{1}]$, and $\Delta = 2.87$ GHz is the zero field splitting of the NV centre electron spin-1's (denoted by $\boldsymbol{S}_A$ and $\boldsymbol{S}_B$). Each NV centre contains a $^{15}$N nuclear spin-1/2 (denoted by $\boldsymbol{I}_A$ and $\boldsymbol{I}_B$, respectively) with an isotropic hyperfine coupling constant $a_N = 3.05$ MHz.

The applied magnetic field $\boldsymbol{B}$ is aligned with NV A. Thus the magnetic quantum number $m_S^A$ is a good quantum number. The Eigenstates of Hamiltonian (2) are denoted by $|\pm\rangle_A$ and $|0\rangle_A$. In this basis, the Hamiltonian is rewritten as

$$H_A = \left(\omega_A^{(+)} + \boldsymbol{h}_A^{(+)}\boldsymbol{I}_A\right)|+\rangle_A\langle+| + \left(\omega_A^{(-)} + \boldsymbol{h}_A^{(-)}\boldsymbol{I}_A\right)|-\rangle_A\langle-| + \omega_A^{(0)}|0\rangle_A\langle 0| \tag{4}$$

where $\boldsymbol{h}_A^{(\pm)} = \pm a_N \hat{z}$ is the effective hyperfine field seen by the $^{15}$N nuclear spin conditioned on the electron spin $|\pm\rangle$ states. For the electron spin in the $|0\rangle$ state, the effective hyperfine field vanishes because ${}_A\langle 0|\boldsymbol{S}_A|0\rangle_A = 0$. Notice that the electron nuclear spin flipflop terms have been neglected in Eq. (4), since the zero field splitting is much larger than the hyperfine coupling strength (i.e., $\Delta \gg a_N$).

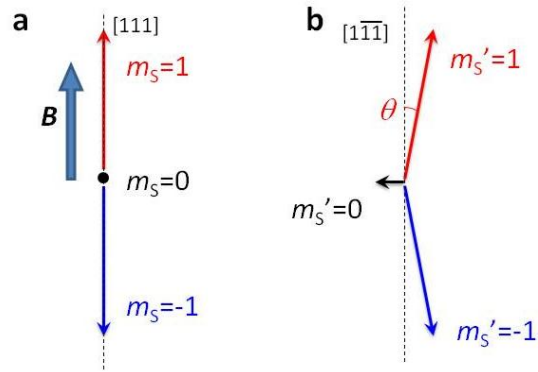

**Supplementary Figure 4 |** **a,** Spin vectors for NV A along [111] direction. The magnetic field $B = 40$ Gauss (the thick blue arrow) is aligned with the NV axis. In this case, the magnetic quantum number $m_S$ is good quantum number. **b,** Spin vectors for NV B in the same field as in a. The spin vectors deviate from the NV axis (i.e. the $[1\bar{1}\bar{1}]$ direction) due to the state mixing effect in a misaligned magnetic field. However, the tilt angle is small because of the weak applied field. For magnetic field $B = 40$ Gauss, the spin magnitudes are +0.998, 0.0001, and -0.998 and the tilt angle $\theta = 1°$. Thus, the states are well-approximately denoted by the magnetic quantum number $m_S = 0$, and $\pm 1$.

Misalignment of the magnetic field to the symmetry axis of NV B causes state mixing. However, for the magnetic field strength ($B \approx 40$ Gauss $\ll |\Delta/\gamma_e|$) applied in this experiments, the effect is small. We can still use the quantum number $m_S^B$ to (approximately) denote the Eigenstates of Hamiltonian (3) (i.e. $|\pm\rangle_B$ and $|0\rangle_B$). Details of the state mixing effect are presented in Supplementary Fig. 4. Similar to Eq. (4) with the electron nuclear spin flipflop terms and the offaxial magnetic field neglected, Hamiltonian (3) is written as

$$H_B = \left(\omega_B^{(+)} + \boldsymbol{h}_B^{(+)}\boldsymbol{I}_B\right)|+\rangle_B\langle+| + \left(\omega_B^{(-)} + \boldsymbol{h}_B^{(-)}\boldsymbol{I}_B\right)|-\rangle_B\langle-| + \left(\omega_B^{(0)} + \boldsymbol{h}_B^{(0)}\boldsymbol{I}_B\right)|0\rangle_B\langle 0|. \tag{5}$$

Please notice that electron spin state $m_S^B = 0$ has now a small magnetic moment and thus a hyperfine interaction $h_B^{(0)}$.

The two NV centres are coupled by the magnetic dipolar interaction given by

$$H_{dip} = \frac{\mu_0}{4\pi} \frac{\gamma_e^2}{r_{AB}^3} [\mathbf{S}_A \cdot \mathbf{S}_B - 3(\mathbf{S}_A \cdot \mathbf{n}_{AB})(\mathbf{n}_{AB} \cdot \mathbf{S}_B)], \qquad (6)$$

where $\mathbf{r}_{AB} = r_{AB}\mathbf{n}_{AB}$ is the displacement vector pointing from NV A to NV B. In general, dipolar interaction between two spins causes the coherence and/or population transfer due to the spin flip-flop process. However, in our experiments, the dipolar coupling (~10 kHz) is much weaker than the energy difference (~10 MHz) of the two NVs. Accordingly, the spin flip-flop terms can be neglected, leaving only the energy shift term as

$$H_{dip} = h\nu_{dip} S_A^z S_B^z, \qquad (7)$$

where $h\nu_{dip}$ is the dipolar coupling strength determined by the DEER measurement. Please note that terms like $S_A^x S_B^z$ can be neglected due to the strong alignment of spin NV A along its z-axis because of the zerofield splitting and the magnetic field. Although the mentioned state mixing in NV B would allow for small contributions from terms like $S_A^z S_B^x$ the experiment shows that for the given relative position these terms are negligible.

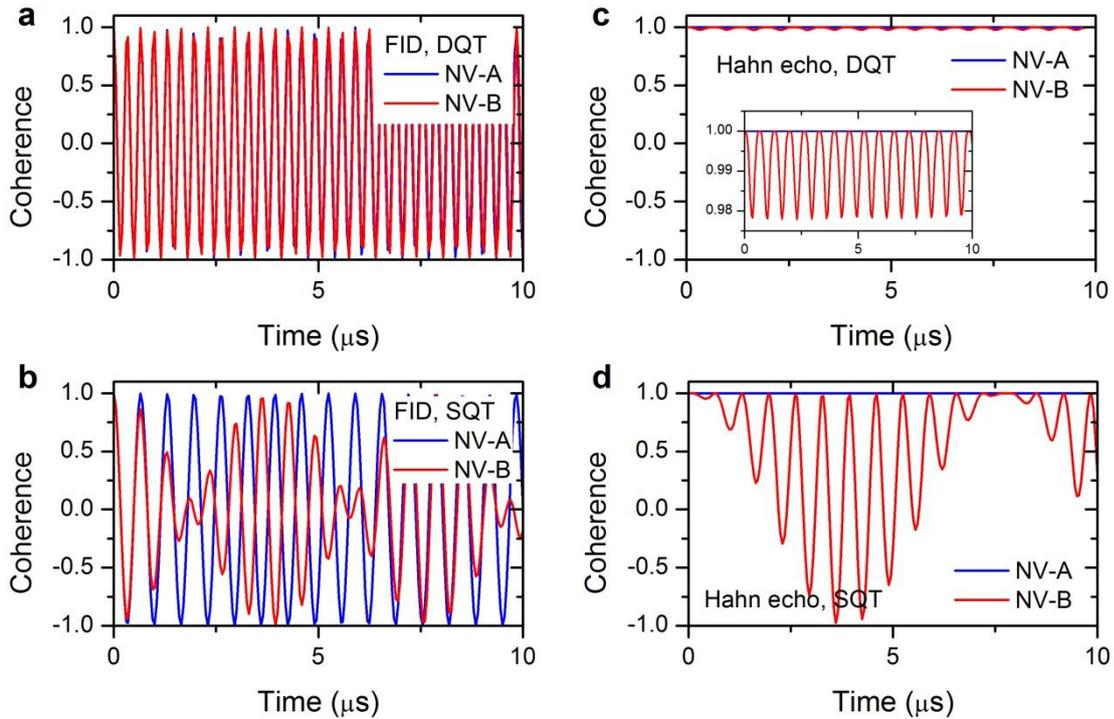

**Supplementary Figure 5 | a, FID modulation due to $^{15}$N nuclear spin.** Upper panel shows the $^{15}$N induced modulation of double quantum transition (DQT), FID for aligned (NV A) and misaligned (NV B) NV centres. The two NV centres have a similar modulation effect. The lower panel shows the

same modulation but for single quantum transition for the two NV centres. **b, Hahn echo modulation due to $^{15}$N nuclear spin**. Upper panel shows the $^{15}$N induced modulation of double quantum transition (DQT) Hahn echo signal for aligned (NV A) and misaligned (NV B) NV centres. The aligned NV centre does not have modulations, and the misaligned NV centre has a modulation with negligible amplitude. Lower panel shows the same modulation effect but for single quantum transition for the two NV centres.

## Single Spin Decoherence

In addition to coupling to the $^{15}$N nuclear spin, the dynamics of the NV centre electron spin is also influenced by the random noise field. The noise field has various sources. In our $^{13}$C isotope purified diamond sample, the residual $^{13}$C nuclear spins provides a *quasi-static* noise. While the motion of electron spins of surrounding defects (e.g., the P1 centre or the defects produced during the ion implantation process) or fluctuating charges impose a *dynamic* noise field on the NV centre electron spin. Such noise is modeled by fluctuating magnetic fields which couple to the NV centre spin levels. Furthermore, we focus on *phase* noise, since the spin relaxation time is much longer than the characteristic time of the entanglement creation process. Including the random magnetic fields $b_A(t)$ and $b_B(t)$, Hamiltonians (4) and (5) become

$$H_k = \left(\omega_k^{(+)} + \boldsymbol{h}_k^{(+)}\boldsymbol{I}_k + b_k(t)\right)|+\rangle_k\langle+| + \left(\omega_k^{(-)} + \boldsymbol{h}_k^{(-)}\boldsymbol{I}_k - b_k(t)\right)|-\rangle_k\langle-| + \left(\omega_k^{(0)} + \boldsymbol{h}_k^{(0)}\boldsymbol{I}_k\right)|0\rangle_k\langle0| \equiv \sum_{m_S=0,\pm} H_k^{(m_S)}|m_S\rangle_k\langle m_S|, \quad (8)$$

where $k$ = A or B and $H_k^{(m_S)}$ is the conditional Hamiltonian for NV centre electron spin states $|m_S\rangle_k$, respectively. The noise fields $b_A(t)$ and $b_B(t)$ cause the NV centre electron spin decoherence. As the measured $T_2$ coherence times of NV A and NV B are significantly different, it is reasonable to assume the noise fields $b_A(t)$ and $b_B(t)$ are independent. In the following, we describe the decoherence process of single NV centres.

### Free-induction decay

For the transition $|m_S\rangle \leftrightarrow |\widetilde{m}_S\rangle$ of NV-$k$, the free-induction decay (FID) of the coherence is calculated as

$$L_{k,FID}^{m_S,\widetilde{m}_S}(t) = \langle \text{Tr}\left[e^{-iH_k^{(m_S)}t}e^{iH_k^{(\widetilde{m}_S)}t}\right]\rangle, \quad (9)$$

where the trace Tr[…] is performed over the $^{15}$N nuclear spin degrees of freedom, and $\langle … \rangle$ is the ensemble average over different realizations of $b_k(t)$ (noise). The $^{15}$N nuclear spins induced dynamics can be separated from dynamics due to the noise field $b_k(t)$ as

$$L_{k,FID}^{m_S,\widetilde{m}_S}(t) = \langle e^{-i(m_S-\widetilde{m}_S)\varphi_k(t)}\rangle \times Tr\left[e^{-i\boldsymbol{h}_k^{(m_S)}\boldsymbol{I}_k t}e^{i\boldsymbol{h}_k^{(\widetilde{m}_S)}\boldsymbol{I}_k t}\right] \equiv L_{k,FID}^{m_S,\widetilde{m}_S} \times G_{k,FID}^{m_S,\widetilde{m}_S}, \quad (10)$$

where $\varphi_k(t)$ is the classical random phase of NV-$k$, i.e.

$$\varphi_k(t) = \int_o^t b_k(t')dt'. \tag{11}$$

Averaging over the random phase $\varphi_k(t)$ gives rise to the irreversible coherence decay $L_{k,FID}^{m_S,\widetilde{m}_S}$, while the $^{15}$N nuclear spin induces a modulation $G_{k,FID}^{m_S,\widetilde{m}_S}$ of the coherence.

### Hahn echo

With a refocusing π pulse, the electron spin coherence under Hahn echo control is

$$L_{k,Hahn}^{m_S,\widetilde{m}_S}(2\tau) = \langle Tr\left[e^{-iH_k^{(m_S)}\tau}e^{-iH_k^{(\widetilde{m}_S)}\tau}e^{iH_k^{(m_S)}\tau}e^{iH_k^{(\widetilde{m}_S)}\tau}\right]\rangle \equiv L_{k,Hahn}^{m_S,\widetilde{m}_S}(2\tau) \times G_{k,Hahn}^{m_S,\widetilde{m}_S}(2\tau) \tag{12}$$

where we have defined the random noise $\varphi_k(t)$ induced decoherence function

$$L_{k,Hahn}^{m_S,\widetilde{m}_S}(2\tau) = \langle e^{-i(m_S-\widetilde{m}_S)\varphi_k(t)}\rangle \tag{13}$$

with the random phase $\varphi_k(t)$ under Hahn echo control

$$\varphi_k(t) = \varphi_k^{(1)} - \varphi_k^{(2)} = \int_0^\tau b_k(t')dt' - \int_\tau^{2\tau} b_k(t')dt' \equiv \int_0^{2\tau} b_k(t')f(t')dt', \tag{14}$$

where $f(t)$ is a filter function taking the value of +1 (-1) before (after) the refocusing π pulse at time $\tau$. The modulation function $G_{k,Hahn}^{m_S,\widetilde{m}_S}(t)$ due to $^{15}$N nuclear spin is

$$G_{k,Hahn}^{m_S,\widetilde{m}_S}(2\tau) = Tr\left[e^{-i\boldsymbol{h}_k^{(m_S)}\boldsymbol{I}_k\tau}e^{-i\boldsymbol{h}_k^{(\widetilde{m}_S)}\boldsymbol{I}_k\tau}e^{i\boldsymbol{h}_k^{(m_S)}\boldsymbol{I}_k\tau}e^{i\boldsymbol{h}_k^{(\widetilde{m}_S)}\boldsymbol{I}_k\tau}\right]. \tag{15}$$

### Modulation effect due to $^{15}$N nuclear spin

The coherence modulation functions due to the 15N nuclear spin is calculated as

$$G_{k,FID}^{m_S,\widetilde{m}_S}(t) = \cos\left(\frac{h_k^{(m_S)}t}{2}\right)\cos\left(\frac{h_k^{(\widetilde{m}_S)}t}{2}\right) + \sin\left(\frac{h_k^{(m_S)}t}{2}\right)\sin\left(\frac{h_k^{(\widetilde{m}_S)}t}{2}\right) \tag{16}$$

and

$$G_{k,Hahn}^{m_S,\widetilde{m}_S}(2\tau) = 1 - 2\left|\sin\left(\frac{h_k^{(m_S)}\tau}{2}\right) \times \sin\left(\frac{h_k^{(\widetilde{m}_S)}\tau}{2}\right)\right|^2. \tag{17}$$

The typical modulation behaviour of a FID and Hahn echo measurement is shown in Supplementary Fig. 5. The direction of the effective hyperfine fields $\boldsymbol{h}_k^{(m_S)}$ and $\boldsymbol{h}_k^{(\widetilde{m}_S)}$ strongly affects the amplitude of the modulation. In particular, for double quantum transitions (i.e., $\Delta m_S = m_S - \widetilde{m}_S = 2$), $\boldsymbol{h}_k^{(+)}$ and $\boldsymbol{h}_k^{(-)}$ are (anti-)parallel with each other (for the misaligned case of NV B, this condition holds approximately). Accordingly, the modulation is negligible (see Supplementary Fig. 5). Therefore the modulation is given by the decoherence induced by the noise field $b_k(t)$

$$L_{k,Hahn}^{+,-}(2\tau) \approx \langle e^{-2i\varphi_k(t)}\rangle = L_k(t). \tag{18}$$

Notice that, for the double quantum transition, the random phase is twice as large as that in the single quantum transition case, while the dipolar coupling strength is enhanced by a factor of 4. As a result, the double quantum transition is used to generate electron spin entanglement more efficiently. Decoherence of individual electron spins $L_A(t)$ and $L_B(t)$ is measured independently and, in the following, we describe the noise effect during the entanglement creation process in terms of the individual spin decoherence functions $L_A(t)$ and $L_B(t)$ without involving any other adjustable parameters.

## Double Electron Electron Resonance (DEER) Experiment

In the DEER experiment, NV A is used to sense the spin state of NV B. By flipping the spin state of NV B, a phase difference scaling with the dipolar interaction strength $\nu_{dip}$ is observable. The electron spin of NV A acquires the same random phase as in Eq. (14), together with an additional phase due to the dipolar coupling to the NV B. Thus, with the $^{15}$N induced modulation neglected, the DEER coherence is expressed as

$$L_{A,DEER}^{+,-}(\tau, t) \approx \langle e^{-2i\varphi_k(t) - i\phi_J(\tau)} \rangle = L_A(t) e^{-i\phi_J(\tau)}, \qquad (19)$$

where $\phi_J(\tau) = 2\pi \nu_{dip} \tau$. Therefore the measured fluorescence intensity is

$$I_{DEER}^A(\tau, t) = L_A \cos(2\pi \nu_{dip} \tau). \qquad (20)$$

The dipolar coupling strength is obtained from the oscillation period for varying $\tau$.

# Entanglement generation and tomography

In this part entanglement generation and tomography is explained in more detail. During the run of the experiments different kinds of entangled states have been prepared. The most basic one includes entanglement between both NV centres in their $m_S = 0$ and $+1$ states $|\Phi_{0+}\rangle = |00\rangle - i|++\rangle$. Entanglement generation is explained on this example as well as its effect on the fluorescence level during generation. Furthermore the generation of the $|\Phi_{+-}\rangle = |++\rangle - i|--\rangle$ entangled state is shown where entanglement is created among the $m_S = -1$ and $+1$ levels of each NV. On this example density matrix tomography will be explained.

## Generation of $|\Phi_{0+}\rangle = |00\rangle - i|++\rangle$ state

### Density matrix evolution

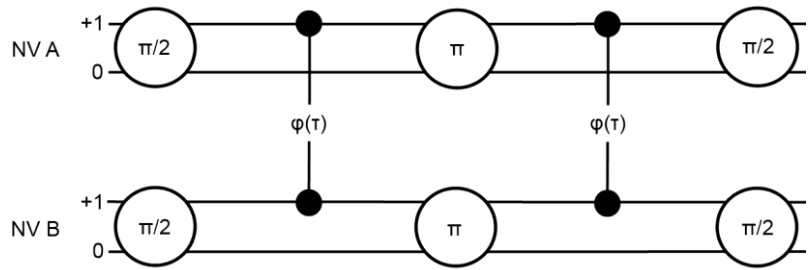

**Supplementary Figure 6 |** Pulse sequence used to create entanglement among electron spin states $m_S$=0 and +1 of both NV centres. A Hahn echo sequence refocuses all quasi static noise. The influence of the NV spins upon each other by magnetic dipolar coupling is inverted by the central $\pi$ pulse. Therefore the interaction is not static and is not refocused but leads to a conditional phase shift gate.

According to the pulse sequence as described in Supplementary Fig. 6, the evolution operator of the entanglement gate $U_{ent}^{|\Phi_{0+}\rangle}$ is

$$U_{ent}^{|\Phi_{0+}\rangle}(\tau) = U_3 U_{evo}(2\tau,\tau) U_2 U_{evo}(\tau,0) U_1, \tag{21}$$

where $U_{1,2,3}$ are the pulse operators as follows (defined in the bases of $\{|+\rangle_k, |0\rangle_k, |-\rangle_k\}$ for $k = A$ or $B$)

$$U_1 = \begin{pmatrix} \frac{1}{\sqrt{2}} & -\frac{1}{\sqrt{2}} & 0 \\ 0 & 0 & -1 \\ \frac{1}{\sqrt{2}} & \frac{1}{\sqrt{2}} & 0 \end{pmatrix}_A \otimes \begin{pmatrix} \frac{1}{\sqrt{2}} & -\frac{1}{\sqrt{2}} & 0 \\ 0 & 0 & -1 \\ \frac{1}{\sqrt{2}} & \frac{1}{\sqrt{2}} & 0 \end{pmatrix}_B, \tag{22}$$

$$U_2 = \begin{pmatrix} 0 & 0 & 1 \\ 0 & -1 & 0 \\ 1 & 0 & 0 \end{pmatrix}_A \otimes \begin{pmatrix} 0 & 0 & 1 \\ 0 & -1 & 0 \\ 1 & 0 & 0 \end{pmatrix}_B, \tag{23}$$

$$U_3 = \begin{pmatrix} \frac{1}{\sqrt{2}} & 0 & \frac{1}{\sqrt{2}} \\ \frac{1}{\sqrt{2}} & 0 & -\frac{1}{\sqrt{2}} \\ 0 & 1 & 0 \end{pmatrix}_A \otimes \begin{pmatrix} \frac{1}{\sqrt{2}} & 0 & \frac{1}{\sqrt{2}} \\ \frac{1}{\sqrt{2}} & 0 & -\frac{1}{\sqrt{2}} \\ 0 & 1 & 0 \end{pmatrix}_B, \tag{24}$$

where $U_1$ and $U_3$ are π/2 rotations and $U_2$ is a π rotation, and all the pulses are applied with the same phase (defined as y-phase, i.e., rotating the electron spin about the y axis in the rotating frame). The evolution operator $U_{evo}(t_1, t_2)$ is defined as

$$U_{evo}(t_1, t_2) = U_J(t_2 - t_1) U_R(t_2, t_1). \tag{25}$$

The operators $U_J$ and $U_R$, respectively, describe the phases accumulation due to the dipolar coupling and the noise field as

$$U_J(\tau) = \text{diag}[e^{-i\phi_J}, 1, e^{i\phi_J}, 1, 1, 1, e^{i\phi_J}, 1, e^{-i\phi_J}], \tag{26}$$

$$U_R(t_2, t_1) = \text{diag}\left[e^{-i\left(\varphi_A^{(i)}+\varphi_B^{(i)}\right)}, e^{-i\varphi_A^{(i)}}, e^{-i\left(\varphi_A^{(i)}-\varphi_B^{(i)}\right)}, e^{-i\varphi_B^{(i)}}, 1, e^{i\varphi_B^{(i)}}, e^{-i\left(-\varphi_A^{(i)}+\varphi_B^{(i)}\right)}, e^{i\varphi_A^{(i)}}, e^{i\left(\varphi_A^{(i)}-\varphi_B^{(i)}\right)}\right],$$

where $\varphi_A^{(i)}$ and $\varphi_B^{(i)}$ are the random phases for NV A and NV B respectively during the $i$th interval ($i$ = 1 or 2, see Eq. 14).

With the evolution operator $U_{ent}^{|\Phi_{0+}\rangle}(\tau)$, the density matrix at time $\tau$ is

$$\rho(\tau) = \langle U_{ent}^{|\Phi_{0+}\rangle}(\tau) \rho_0 \left(U_{ent}^{|\Phi_{0+}\rangle}(\tau)\right)^\dagger \rangle, \tag{27}$$

where $\langle ... \rangle$ denotes the ensemble averaging over different random noise realizations, and $\rho_0 = |00\rangle\langle 00|$ is the initial density matrix. Straightforward calculation gives

$$\rho_{|\Phi_{0+}\rangle}(\tau) = \begin{pmatrix} \langle|Q_4|^2\rangle & \langle Q_2^*Q_4\rangle & 0 & \langle Q_1^*Q_4\rangle & \langle Q_3^*Q_4\rangle & 0 & 0 & 0 & 0 \\ \langle Q_4^*Q_2\rangle & \langle|Q_2|^2\rangle & 0 & \langle Q_1^*Q_2\rangle & \langle Q_3^*Q_2\rangle & 0 & 0 & 0 & 0 \\ 0 & 0 & 0 & 0 & 0 & 0 & 0 & 0 & 0 \\ \langle Q_4^*Q_1\rangle & \langle Q_2^*Q_1\rangle & 0 & \langle|Q_1|^2\rangle & \langle Q_3^*Q_1\rangle & 0 & 0 & 0 & 0 \\ \langle Q_4^*Q_3\rangle & \langle Q_2^*Q_3\rangle & 0 & \langle Q_1^*Q_3\rangle & \langle|Q_3|^2\rangle & 0 & 0 & 0 & 0 \\ 0 & 0 & 0 & 0 & 0 & 0 & 0 & 0 & 0 \\ 0 & 0 & 0 & 0 & 0 & 0 & 0 & 0 & 0 \\ 0 & 0 & 0 & 0 & 0 & 0 & 0 & 0 & 0 \\ 0 & 0 & 0 & 0 & 0 & 0 & 0 & 0 & 0 \end{pmatrix}. \quad (28)$$

The non-zero ensemble average values of $\langle Q_i^*Q_j\rangle$ for $i = 1 \ldots 4$ are calculated as

$$\langle|Q_1|^2\rangle = \tfrac{1}{4}\big[1 - L_A L_B + (L_A - L_B)\cos(4\phi_J)\big], \quad (29)$$

$$\langle|Q_2|^2\rangle = \tfrac{1}{4}\big[1 - L_A L_B - (L_A - L_B)\cos(4\phi_J)\big], \quad (30)$$

$$\langle|Q_3|^2\rangle = \tfrac{1}{4}\big[1 - L_A L_B + (L_A + L_B)\cos(4\phi_J)\big], \quad (31)$$

$$\langle|Q_4|^2\rangle = \tfrac{1}{4}\big[1 - L_A L_B - (L_A + L_B)\cos(4\phi_J)\big], \quad (32)$$

$$\langle Q_1^*Q_2\rangle = \langle Q_2^*Q_1\rangle^* = -\tfrac{i}{4}(L_A - L_B)\sin(4\phi_J), \quad (33)$$

$$\langle Q_3^*Q_4\rangle = \langle Q_4^*Q_3\rangle^* = -\tfrac{i}{4}(L_A + L_B)\sin(4\phi_J). \quad (34)$$

At time $\tau^* = 1/(16\nu_{dip})$, the density matrix reads

$$\rho_{|\Phi_{0+}\rangle}(\tau^*) = \frac{1}{4}\begin{pmatrix} 1+\pi_{AB} & 0 & 0 & 0 & -i\sigma_{AB} & 0 & 0 & 0 & 0 \\ 0 & 1-\pi_{AB} & 0 & -i\delta_{AB} & 0 & 0 & 0 & 0 & 0 \\ 0 & 0 & 0 & 0 & 0 & 0 & 0 & 0 & 0 \\ 0 & i\delta_{AB} & 0 & 1-\pi_{AB} & 0 & 0 & 0 & 0 & 0 \\ i\sigma_{AB} & 0 & 0 & 0 & 1+\pi_{AB} & 0 & 0 & 0 & 0 \\ 0 & 0 & 0 & 0 & 0 & 0 & 0 & 0 & 0 \\ 0 & 0 & 0 & 0 & 0 & 0 & 0 & 0 & 0 \\ 0 & 0 & 0 & 0 & 0 & 0 & 0 & 0 & 0 \\ 0 & 0 & 0 & 0 & 0 & 0 & 0 & 0 & 0 \end{pmatrix}, \quad (35)$$

where $\pi_{AB} = L_A L_B$, $\delta_{AB} = L_A - L_B$ and $\sigma_{AB} = L_A + L_B$. In the idea case (i.e., without decoherence $L_A = L_B = 1$), Eq. (35) gives the desired entangled state $|\Phi_{0+}\rangle = |00\rangle - i|++\rangle$. Taking into account the decoherence effect, the fidelity of the entanglement gate is calculated as

$$F(\tau) = Tr[\rho(\tau)|\Phi_{0+}\rangle\langle\Phi_{0+}|] = \tfrac{1}{4}\big[1 + L_A L_B + (L_A + L_B)\sin(8\pi\nu_{dip}\tau)\big]. \quad (36)$$

**Fluorescence intensity**

The observed fluorescence intensity, which is related to the zero-state population $P_0$, is expressed as

$$\langle P_0\rangle = \langle Tr\left[\hat{P}_M U_{ent}^{|\Phi_{0+}\rangle}\rho_0\left(U_{ent}^{|\Phi_{0+}\rangle}\right)^\dagger\right]\rangle, \quad (37)$$

where the optical readout process is described by the measurement operator $\hat{P}_M = \alpha|0\rangle_A\langle 0| + \beta|0\rangle_B\langle 0|$ with $\alpha$ and $\beta$ being the fluorescence coefficients of NV A and NV B,

respectively. With the evolution operator described in previous section, the fluorescence intensity at time $\tau$ is

$$\langle P_0(\tau)\rangle = \frac{\alpha+\beta}{2} + \frac{\tilde{\alpha}+\tilde{\beta}}{2}\cos(8\pi\nu_{dip}\tau), \qquad (37)$$

where $\tilde{\alpha} = \alpha \cdot L_A(t)$ and $\tilde{\beta} = \beta \cdot L_A(t)$ are the corrected florescence coefficients by the decoherence effect.

## Generation of $|\Phi_{+-}\rangle = |++\rangle - i|--\rangle$ state

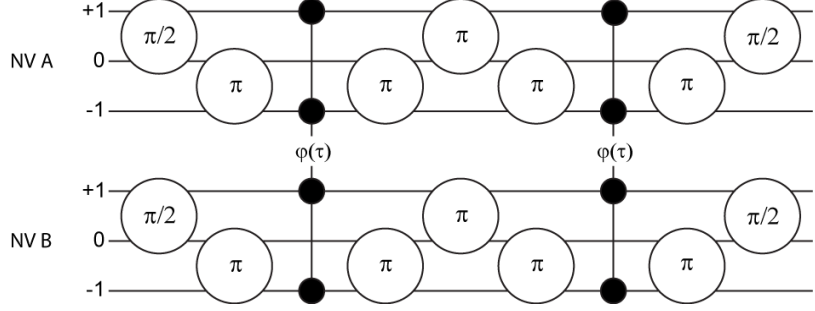

**Supplementary Figure 7 |** Pulse sequence used to create the entangled state. In contrast to the sequence depicted in Supplementary Fig. 6 here the contributing spin states are $m_S = -1$ and $+1$.

Starting from the entangled state $|\Phi_{0+}\rangle = |00\rangle - i|++\rangle$, we can generate other entangled states by applying pulses locally manipulating the two NV centres. For example by applying $\pi$ pulses on the $|0\rangle \leftrightarrow |-\rangle$ transitions of NV A and NV B respectively, we convert $|\Phi_{0+}\rangle$ into the $|\Phi_{+-}\rangle = |++\rangle - i|--\rangle$ state.

The evolution operator for $|\Phi_{+-}\rangle$ state generation process is $U_{ent}^{|\Phi_{+-}\rangle} = \pi_{0-}^{(A)}\pi_{0-}^{(B)}U_{ent}^{|\Phi_{0+}\rangle}$ with $\pi_{0-}^{(A/B)}$ being the $\pi$ rotations on NV A/B. The florescence intensity $P_g(t)$ in the phase measurement is expressed as

$$P_g(t) = \langle Tr\left[\hat{P}_M U_{back} U_{evo(t)} U_{ent}^{|\Phi_{+-}\rangle} \rho_0 \left(U_{ent}^{|\Phi_{+-}\rangle}\right)^\dagger U_{evo}^\dagger(t) U_{back}^\dagger\right]\rangle, \qquad (38)$$

where $U_{back}^\dagger = \left(U_{ent}^{|\Phi_{+-}\rangle}\right)^\dagger$ is the inverse evolution operator. When we drive the NV centres at the microwave frequencies which are symmetrically detuned from the hyperfine levels, the phase measurements exhibit the oscillation with frequency $\delta_{AB} = 6$ MHz as

$$P_g(t) = \frac{\alpha+\beta}{2} + \frac{\alpha L_A^2 + \beta L_B^2}{2}[1 + \cos(2\pi\delta_{AB}t)]. \qquad (39)$$

In addition to the above explained method to generate entangled state $|\Phi_{+-}\rangle = |++\rangle - i|--\rangle$ we can also exploit the spin 1 nature of the electron spin to speed up the entanglement sequence by a factor of four. Therefore we create local superposition states on the $m_S = -1$ and $+1$ levels of the individual NV centres during the entanglement sequence (see Supplementary Fig. 7).

# Collective phase evolution of entangled states

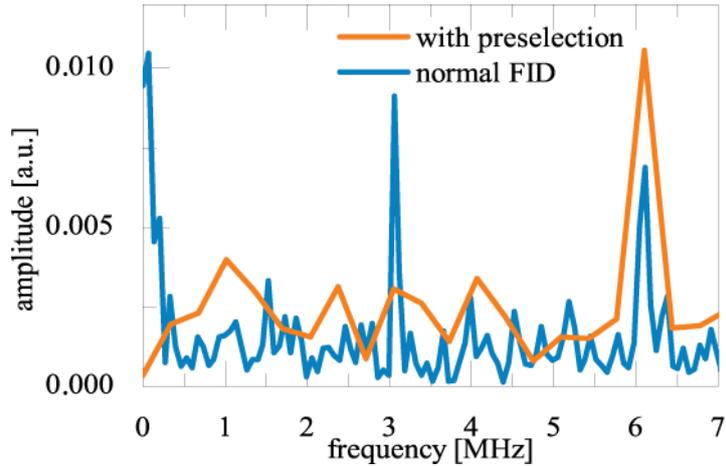

**Supplementary Figure 8 |** Fourier transformation of the phase measurements of the $\Phi_{DQ}^+ = \frac{1}{\sqrt{2}}(|-1-1\rangle + i|11\rangle)$ state where each microwave frequency is detuned by $\pm 1.5$ MHz. The blue curve is a measurement without charge state detection. The smaller second peak at 3 MHz can be attributed to a single NV double quantum evolution due to the other NV being in the neutral charge state. The peak at 6 MHz is due to the collective phase accumulation of the entangled state. The blue line is the same measurement but with charge state pre-selection where only the collective phase peak is observed. The 3 MHz peak vanishes as expected and the low frequency peak vanishes due to base line reduction for the FFT.

A fundamental property of entangled states is the evolution of their collective phases[14]. Due to the inseparable wave function both entities do not evolve independently of each other but evolve as one. Therefore the phase collected by the system is not the single entity phases but the sum (for $\Phi$-like entangled state) or the difference (for $\Psi$-like entangled state) of the single entity phases. With a phase evolution measurement it is possible to probe this fundamental property. In Supplementary Fig. 9 the FFT of such an experiment is shown. Here coherences between $m_S = +1$ and $m_S = -1$ were used with a single entity evolution frequency of $\pm 3$ MHz. An entangled state therefore yields evolution frequencies of 0 and 6 MHz. As shown in Supplementary Fig. 12 the phase evolution shows a peak at 6 MHZ indicating entanglement, but also a peak at 3 MHz indicating single entity evolution. The latter is due to the charge state of the NV centre and can be overcome by charge state detection as shown by the orange line.

# State tomography

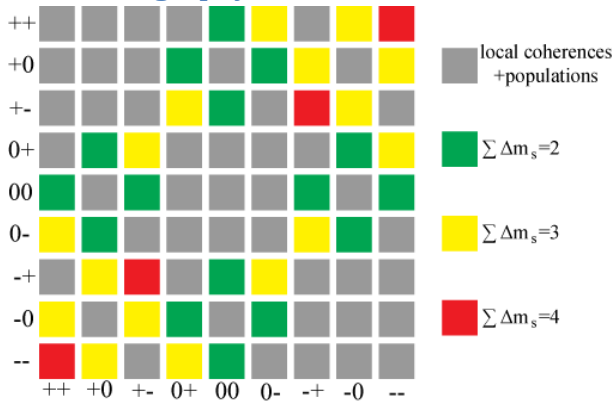
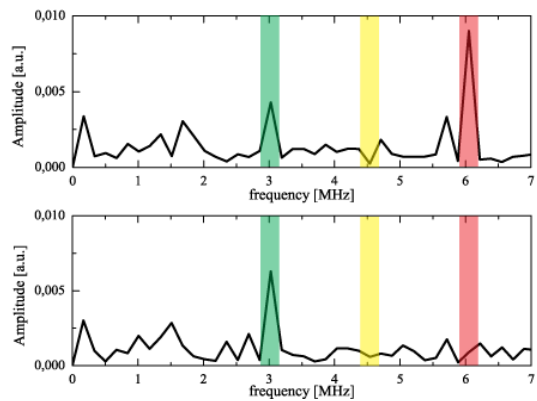

**Supplementary Figure 7 | Evolution of collective phase of entangled state. a,** Density matrix of the NV-NV system color-coded to show the phase evolution frequencies. The NV-NV coherences are

shown in green, yellow and red while the grey fields correspond to single NV coherences and populations. For all measurements the detuning is chosen to be $\pm 1.5$ MHz. The expected collective phase oscillation is given by $\sum \Delta m_S \cdot 1.5$ MHz. **b,** Fast Fourier transformed (FFT) phase evolution measurement of a $\Phi_{DQ}^+ = \frac{1}{\sqrt{2}}(|-1-1\rangle + i|11\rangle)$ entangled state. The peak at 6 MHz corresponds to the phase evolution of the entangled state ($\sum \Delta m_S = 4$). The peak at 3 MHz corresponds to the single spin evolution due to NV$^0$. **c,** FFT of an evolution of desired initial entangled state $\Phi_{DQ}^+ = \frac{1}{\sqrt{2}}(|-1-1\rangle + i|11\rangle)$ altered to $\frac{1}{\sqrt{2}}(|-10\rangle - i|11\rangle)$. The back transformation was carried out using the reverse entanglement gate of the $\Phi_{DQ}^+$ state. The collective phase of the altered state should evolve at 4.5 MHz (yellow line, ($\sum \Delta m_S = 3$) but we are only probing the 6 MHz phase frequency ($\sum \Delta m_S = 4$, red line). As expected for an ideal initial entangled state no peak at 4.5 MHz and 6 MHz is visible. The only peak visible is the 3 MHz peak due to single spin evolution because charge state initialization was not performed.

Due to improper generation of the entangled states, populations and coherences may appear in the density matrix that are different from what is expected. These entries will be revealed by density matrix tomography. By analyzing the evolution of the phases of the generated entangled states the off-diagonal entries of the density matrix can be reconstructed.

Supplementary Fig. 9 shows local coherences and populations as well as the collective phases that evolve faster or slower than local coherences. To check for their presence the generated entangled state (e.g. $\Phi_{DQ}^+ = \frac{1}{\sqrt{2}}(|-1-1\rangle + i|11\rangle)$) is first altered by local quantum gates to shift the target phases onto the $\sum \Delta m_S = 4$ phase (e.g. a π pulse on NV A on the $m_S = 0 \leftrightarrow \rightarrow -1$ transition to check for a possible $\frac{1}{\sqrt{2}}(|-10\rangle \pm i|11\rangle)$ coherence). This altered state then evolves freely and is finally transformed by the reverse entanglement gate to convert the $\sum \Delta m_S = 4$ phase into a population difference. The amplitude at phase frequency $\sum \Delta m_S = 4$ depends on the altered state and allows an estimate of the target phase of the initial entangled state (see Supplementary Fig. 9). The detuning of the microwave transitions was chosen to be $\pm 1.5$ MHz (in the centre of the hyperfine interaction of the $^{15}$N). The measured peak intensity is shown in Supplementary Fig. 10.

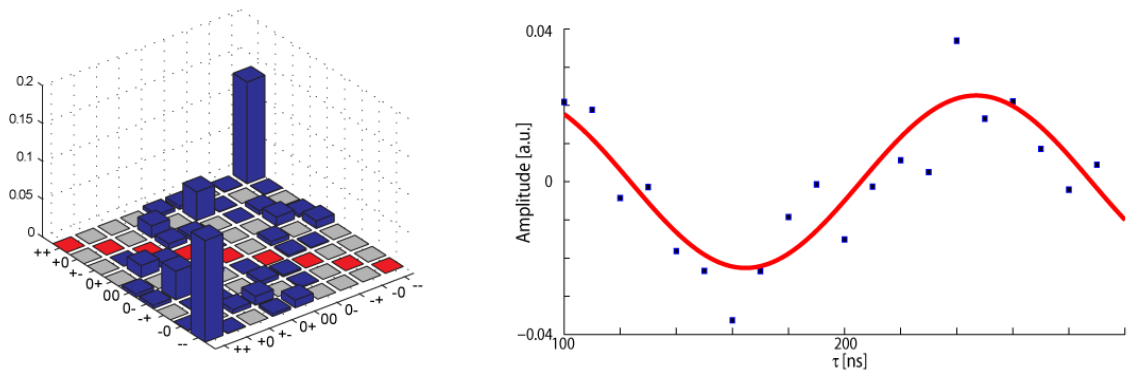

**Supplementary Figure 8 | Density matrix tomography. a,** The blue bars represent the measured oscillation amplitude at the frequency derived from Supplementary Fig. 7. The grey bars are local coherences and the red bars are the populations. **b,** measured collective phase evolution of the $\Phi_{DQ}^+ = \frac{1}{\sqrt{2}}(|-1-1\rangle + i|11\rangle)$ state with charge state pre-selection.

In order to save measurement time, these measurements were performed without charge state pre-selection. To normalize the measurement to a charge state preselected scheme the amplitude of the phase measurement of the $\Phi_{DQ}^+ = \frac{1}{\sqrt{2}}(|-1-1\rangle + i|11\rangle)$ state with charge state pre-selection was correlated to the measurements without charge state pre-selection. Please note that the measured amplitude was only half of the coherence signal, since a detuning of ±1.5 MHz can either result in a 6 MHz or a 0 MHz evolution. The resulting NV-NV coherences are shown in Supplementary Fig. 11 a. The NV-NV correlations close to zero should be regarded as upper limits given by the uncertainty of the measurement.

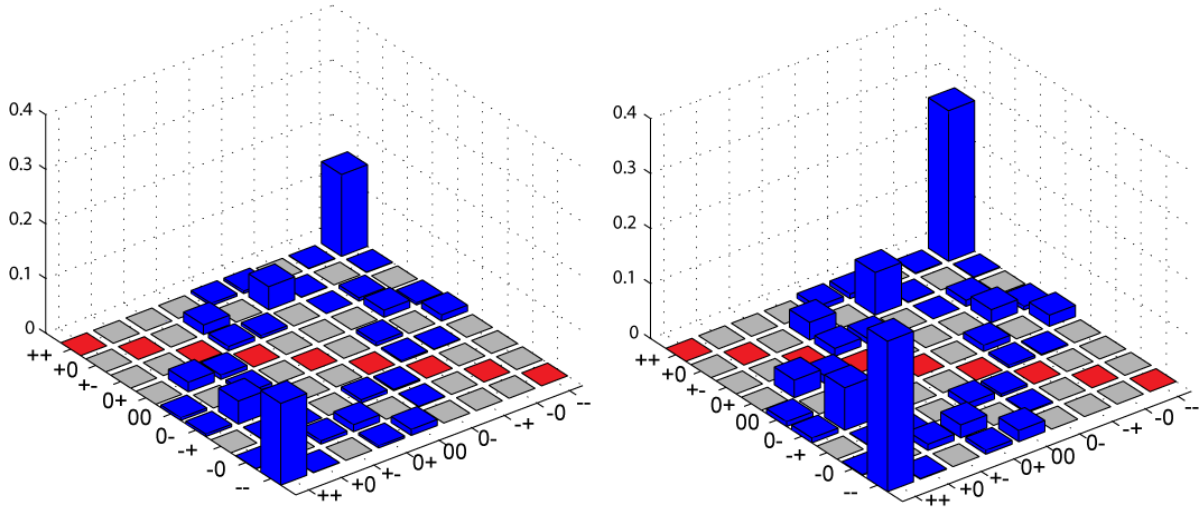

**Supplementary Figure 9 | Density matrices of entangled states. a,** Measured density matrix of the NV-NV coherences normalized with the phase evolution contrast. **b,** NV-NV coherences normalized for the second entanglement gate used to project the coherences into a detectable population difference

Since the entanglement gate is applied twice, once to create the entanglement and once to project the entangled state onto a readable population difference, the measured NV-NV coherences are corrected to represent the actually NV-NV coherences after the first entanglement gate. A NV initialization close to unity is assumed. Supplementary Fig. 11 b shows the NV-NV coherences after the entanglement gate. To probe the local NV coherences they were projected onto population differences using π/2 pulses. The main diagonal was probed using π pulses to collect the uncorrelated population of each NV. With a DEER measurement using the one NV centre as sensor it is possible to weight the NV populations and calculate the main diagonal. In Supplementary Fig. 12 the full reconstructed density matrix is shown. The nonidentical populations of $|11\rangle$ and $|-1-1\rangle$ is most likely due to pulse errors.

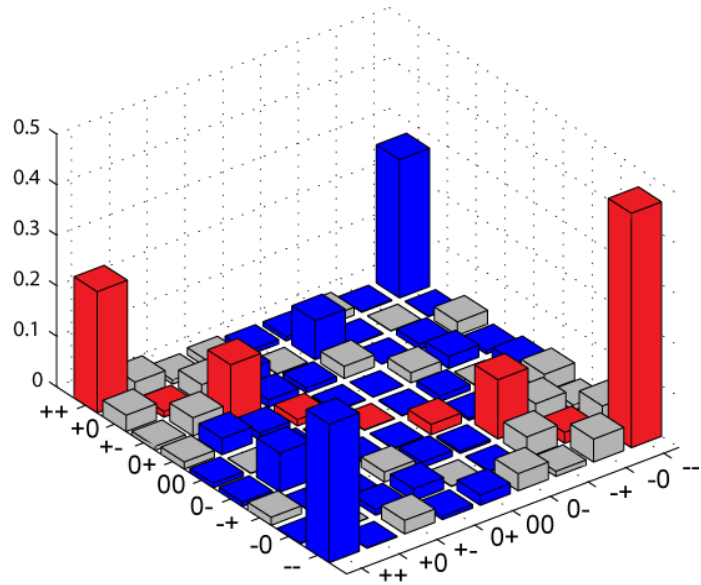

**Supplementary Figure 10 |** Full reconstructed density matrix with the populations in red, the local coherences in grey and the NV-NV coherences in blue.

## Entangled state lifetime

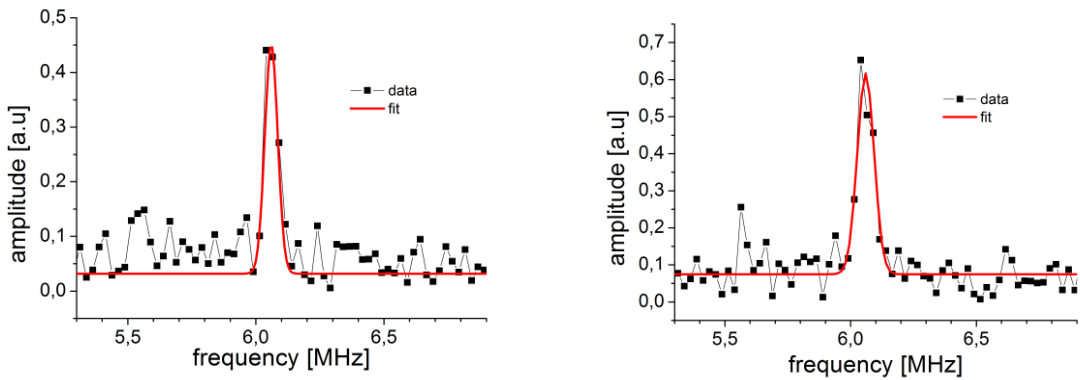

**Supplementary Figure 11 |** Entanglement life time for **a,** $T(\Phi_{DQ}^-)$ and **b,** $T(\Psi_{DQ}^-)$. The frequency spectrum was fitted with a Gaussian.

Free evolution lifetime ($T_2^*$-limited) of the entangled states have been measured by performing phase accumulation measurements on a long time scale. The measured data was Fourier transformed allowing identifying and fitting the entanglement component in the frequency spectrum. To increase the quality of the fit zeros were added. As shown in Supplementary Fig. 13 a Gaussian was fitted yielding a lifetime $T(\Phi_{DQ}^-) = 28.2 \pm 2.2$ µs and $T(\Psi_{DQ}^-) = 23.7 \pm 1.7$ µs. The evolution of the entangled state was measured up to 40 µs allowing deducing that the fit of the frequency spectrum was indeed limited by the lifetime and not due to limitations of the acquisition window.

# Entanglement of remote nuclear spins

For the entanglement of remote nuclear spins (namely the $^{15}$N nuclear spins of each NV of the pair) the entangled state of the electron spins has to be swapped onto the nuclear spins. As these nuclear spins have a mutual interaction that is far less than their lifetimes we like to call these nuclear spins remote. In addition to the appealing effect of having created a really non-local quantum state the swap onto the nuclear spins has also the benefit of exploiting their much longer coherence lifetime which protects the entangled state.

To realize the swap coherent control of the nuclear spins is necessary. We like to implement this without the aid of radiofrequency radiation which is possible as shown in[15].

## Nuclear spin operations without radio frequency fields

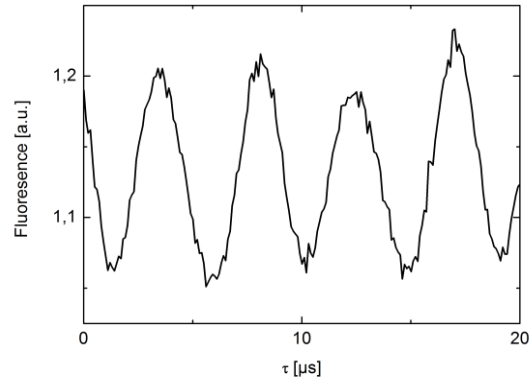

**Supplementary Figure 12 |** CNOT Rabi oscillations according to Supplementary Eq. (41).

In order to transfer the entanglement from the electrons to the $^{15}$N nuclei intrinsic to our NV centres coherent manipulation of the nuclear spin depending on the state of the electron spin is necessary. In principal the application of an RF field at the $^{15}$N resonance frequency allows for such control. However in the setup configuration used for the experiments it was not possible to generate a strong enough RF field at the NV centres. However a combination of a magnetic field perpendicular to the NV centre axis and selective pulses on the electron spin allow for coherent manipulation of the nuclear spin.

The dynamics of the $^{15}$N nuclear spin is governed by the following Hamiltonian conditioned on the electron spin state $m_S$[16]

$$H_{^{15}N} = a_{zz} S_z I_z + \gamma_N \boldsymbol{B} \cdot \underline{\underline{g}}(m_S) \cdot \boldsymbol{I} \qquad (40)$$

where $\underline{\underline{g}}(m_s)$ is the effective g-factor tensor

$$\underline{\underline{g}}(m_s) = \begin{pmatrix} 1 & 0 & 0 \\ 0 & 1 & 0 \\ 0 & 0 & 1 \end{pmatrix} - \frac{\gamma_e}{\gamma_N \Delta}(2 - 3|m_s|)\begin{pmatrix} a_{xx} & a_{xy} & a_{xz} \\ a_{yx} & a_{yy} & a_{yz} \\ 0 & 0 & 0 \end{pmatrix}$$

where $a_{ij}$ is the hyperfine tensor component. In the $^{15}$N nuclear spin case $a_{xx} = a_{yy} = a_{zz} = a_N = 3.05$ MHz and $a_{i \neq j} = 0$. The second term of the g-factor tensor describes the enhancement effect due to the electron spin state mixing.

For electron spin in $m_s = \pm 1$ states, the $^{15}$N nuclear spin is well quantized along the NV-axis due to the strong hyperfine coupling to the electron spin (the first term in Eq. 40). While for electron spin in $m_s = 0$ state, the direct hyperfine coupling vanishes, and the $^{15}$N nuclear spin will precess about an effective field $\gamma_N \boldsymbol{B} \cdot \underline{\underline{g}}(m_s = 0)$. Without loss of generality, we assume the magnetic field is applied in the x-z plane with a polar angle $\theta$ with respect to the z-axis (NV axis). In this case, the $^{15}$N nuclear spin Hamiltonian reads (in the basis of $|\pm 1/2\rangle$ pointing along z-axis)

$$H_{^{15}N}(m_s = 0) = \frac{1}{2}\begin{pmatrix} \omega_z & \omega_x \\ \omega_x & -\omega_z \end{pmatrix}$$

with $\omega_x = \gamma_N B \sin(\theta)(1 - \frac{2\gamma_e}{\gamma_N}\frac{a_N}{\Delta})$ and $\omega_z = \gamma_N B \cos(\theta)$. In ideal case where the angle $\theta$ approaches to $\pi/2$, the nuclear spin will have a perfect Rabi oscillation about x-axis. In the realistic case, as long as the condition $|\omega_x| \gg |\omega_z|$ is satisfied, which is the case of our measurements, one will obtain a high-fidelity Rabi oscillation. The population of $|m_s = 0\ m_I = -1/2\rangle$ state is given by

$$\rho_{m_s=0\ m_I=-1/2} = \cos^2\left(\frac{\Omega t}{2}\right) + \frac{\omega_z^2}{\Omega^2}\sin^2\left(\frac{\Omega t}{2}\right) \quad (41)$$

with $\Omega = \sqrt{\omega_x^2 + \omega_z^2}$ as oscillation frequency as shown in Supplementary Fig. 14.

### Entanglement SWAP/storage

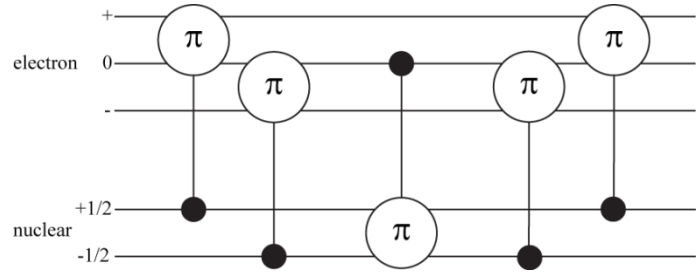

**Supplementary Figure 13 | Entanglement storage sequence.** conditional π rotations were used to store the electron spin entanglement in the nuclear spins. The circle represents a π pulse between the overcast states conditional on the state of the nuclear/electron spin represented by the line and the black dot.

In order to store the electron spin entanglement in the nuclear spins the populations and coherences have to be swapped between the electron and the nuclear spin. This can be achieved with conditional π-pulses (see Supplementary Fig. 15).

Without initializing the nuclear spins, the nuclear spin is in a thermal superposition state given by

$$|++\rangle + |--\rangle \otimes \left|\frac{1}{2};\frac{1}{2}\right\rangle$$

$$|++\rangle + |--\rangle \otimes \left|\frac{1}{2};-\frac{1}{2}\right\rangle$$

$$|++\rangle + |--\rangle \otimes \left|-\frac{1}{2};\frac{1}{2}\right\rangle$$

$$|++\rangle + |--\rangle \otimes \left|-\frac{1}{2};-\frac{1}{2}\right\rangle$$

As a first step a π pulse on the $m_S = 0 \leftrightarrow +1$ transition conditional on the nuclear spin state $m_I = +1/2$ and a π pulse on $m_S = 0 \leftrightarrow -1$ transition conditional on $m_I = -1/2$ were applied on both electron spins simultaneously

$$|00\rangle + |--\rangle \otimes |1/2; 1/2\rangle$$

$$|0+\rangle + |-0\rangle \otimes |1/2; -1/2\rangle$$

$$|+0\rangle + |0-\rangle \otimes |-1/2; 1/2\rangle$$

$$|++\rangle + |00\rangle \otimes |-1/2; -1/2\rangle$$

Then a nuclear spin π rotation ($m_I = +1/2 \leftrightarrow -1/2$) conditional on the electron spin state $m_S = 0$ creates a coherent state between the nuclear spins and the electron spins. For this nuclear π rotation orthogonal magnetic field components were used as described in the previous section.

$$|00\rangle|-1/2; -1/2\rangle + |--\rangle|1/2; 1/2\rangle$$

$$|0+\rangle|-1/2; -1/2\rangle + |-0\rangle|1/2; 1/2\rangle$$

$$|+0\rangle|-1/2; -1/2\rangle + |0-\rangle|1/2; 1/2\rangle$$

$$|++\rangle|-1/2; -1/2\rangle + |00\rangle|1/2; 1/2\rangle$$

In order to store the entanglement completely in the nuclear spin a π pulse on the $m_S = 0 \leftrightarrow +1$ transition conditional on the nuclear spin state $m_I = +1/2$ and a π pulse on $m_S = 0 \leftrightarrow -1$ transition conditional on $m_I = -1/2$ were applied on both electron spins simultaneously

$$|--\rangle|-1/2; -1/2\rangle + |--\rangle|1/2; 1/2\rangle$$

$$|-+\rangle|-1/2; -1/2\rangle + |-+\rangle|1/2; 1/2\rangle$$

$$|+-\rangle|-1/2; -1/2\rangle + |+-\rangle|1/2; 1/2\rangle$$

$$|++\rangle|-1/2; -1/2\rangle + |++\rangle|1/2; 1/2\rangle$$

Now the coherence is fully stored in the nuclei and the electron spin wave function can be separated from the nuclear spin wave function. Please note, that for perfect π pulses the storage efficiency is 1. However due to the non-perfect nuclear spin rotations the storage efficiency is reduced.

The nuclear spin state manipulation used in this letter works best for a 90° angle between the NV axis and magnetic field. However, the electron spin $T_2$ is shortened at non-aligned magnetic fields[17] and additionally the magnetic field interaction is weakened for parallel magnetic field close to zero[18], therefore the fidelity of the entanglement creation becomes rather poor. As a compromise an angle of 54.5° was chosen to demonstrate the entanglement storage leading to a reduced storage efficiency.

# Photon correlations

With fluorescence measurements one can only determine the mean population of both NVs. However no information about the population correlation is determinable. With two photon correlations the population correlation of an unknown state ρ is determinable. The measurement is conducted in Hanbury-Brown-Twiss configuration of two photo detectors one sending the start signal and the other the stop signal of the measurement. By recording the start events and stop events it is possible to calculate the two photon emission probability in relation to the one photon events. Assuming identical photon emission rates for the two defects the two photon emission probability for different states $\rho = |\phi\rangle\langle\phi|$ is given by

$$\rho_{00} = \frac{k_0 k_0}{k_0 + k_0} = \frac{k_0}{2}$$

$$\rho_{11} = \frac{k_1 k_1}{k_1 + k_1} = \frac{k_1}{2}$$

$$\rho_{01} = \rho_{10} = \frac{k_0 k_1}{k_0 + k_1}$$

Where $k_0$ is the photon emission probability for $m_S = 0$ and $k_1$ is the photon emission probability of $m_S = \pm 1$. The suffixes on ρ indicate the state of the NV centres. For an uncorrelated superposition state $\rho_{uncor}$ the two photon probability is given by

$$\rho_{uncor} = \frac{k_0 + k_1}{4}$$

whereas for a correlated state it is given by

$$\rho_\Phi = \frac{k_0^2 + k_1^2}{2(k_0 + k_1)}$$

$$\rho_\Psi = \frac{k_0 k_1}{k_0 + k_1}$$

Where $\Phi = 1/\sqrt{2}(|00\rangle + |11\rangle)$ and $\Psi = 1/\sqrt{2}(|01\rangle + |10\rangle)$.

Any superposition state with $m_S = 0$ and $+1$ of the two NV spins can be described in a basis $\Phi^\pm = 1/\sqrt{2}(|00\rangle \pm |11\rangle)$ and $\Psi^\pm = 1/\sqrt{2}(|01\rangle \pm |10\rangle)$

$$\phi = \alpha^+ \Phi^+ + \alpha^- \Phi^- + \beta^+ \Psi^+ + \beta^- \Psi^-$$

where $0 \leq |\alpha^\pm|, |\beta^\pm| \leq 1$ and $\sum_{a=\alpha^\pm,\beta^\pm} a^2 = 1$. Photon correlations do not distinguish between $\Phi^+$ and $\Phi^-$ as well as $\Psi^+$ and $\Psi^-$. Therefore we restrict ourselves to $\alpha^2 = (\alpha^+)^2 + (\alpha^-)^2$ and $\beta^2 = (\beta^+)^2 + (\beta^-)^2$ with $\alpha^2 + \beta^2 = 1$. For $\alpha^2 + \beta^2 = 1/2$ ϕ is uncorrelated. The values of $\alpha^2$ and $\beta^2$ can be calculated with

$$\sqrt{\alpha^2} = \frac{2(Sk_0 + Sk_1 - k_0 k_1)}{(k_0 - k_1)^2}$$

$$\sqrt{\beta^2} = \frac{-2Sk_0 + k_0^2 - 2Sk_1 + k_1^2}{(k_0 - k_1)^2}$$

Where $k_0$ is measured signal for $|00\rangle$, $k_1$ is measured signal for $|11\rangle$ and $S$ is the measured signal for state $\phi$.

The fact that an uncorrelated state is supposed to have $\alpha^2 = \beta^2 = 1/2$ was used to determine the error in the measurement.

Using the different exited state life times for the spin states ($\tau_{m_s=0} = 23$ ns, $\tau_{m_s=\pm 1} = 12.7$ ns [19]) the signal can be enhanced even further by only selecting two photon correlations with a delay larger than $\tau_{m_s=\pm 1}$. In Fig. 3 of the main text the time resolved photon correlation measurements are shown and the man diagonal is calculated. Please note that with local π/2 pulses not only the $S_z$ correlations are accessible but also the $S_{x/y}$ correlations and therefore a full spin state tomography is possible.

# Microwave assisted superresolution microscopy

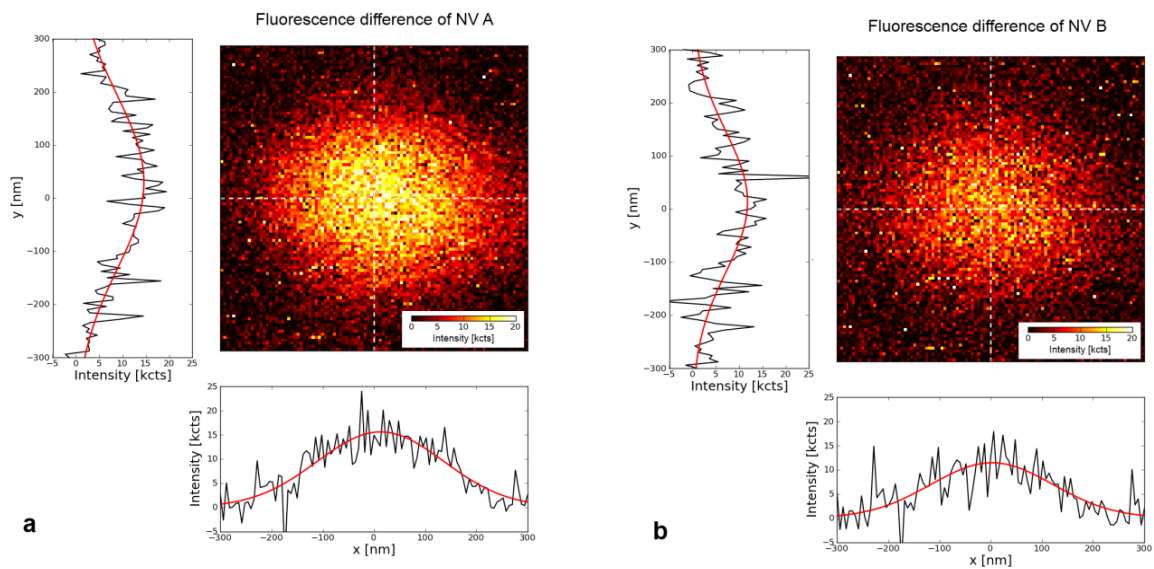

**Supplementary Figure 16 | Superresolution microscopy. a,b,** Fluorescence difference signal of NV A respective B. The images are obtained by subtracting the fluorescence of the NV pair in the state $m_S^A = +1, m_S^A = 0$ ($m_S^A = 0, m_S^B = +1$ for right image) from the reference signal with $m_S^A = m_S^B = 0$. Scans along the lines marked in the 2D images are shown left and below the images.

If the point spread function is known, a single object can be located with an arbitrary precision by fitting the fluorescence image with the point spread function. In order to obtain a fluorescence image of only one emitter on can rely on different techniques, e.g. PALM and STORM. The method used in this work however does not exploit the stochastic blinking behaviour of a photon emitter. Instead the emitter is deterministically switched to a darker state by applying a π-pulse on its ground state. Here the same pulse sequence as used for ODMR spectroscopy is applied (laser (300 ns) – wait (1 μs) – π-pulse). Each point of the confocal scan is then measured three times, once with a π-pulse on NV A and NV B respectively and once without any pulse as reference signal. By subtracting the fluorescence data with the π-pulse from the reference data the fluorescence intensity from a single emitter is obtained (see Supplementary Fig. 16). A fit of the point spread function can be made. However, systematic errors in the position of the fit can arise if the point spread function is not ideally matched by the fit model function. In order to eliminate such errors we calculate the convolution of the two difference signals (see Supplementary Fig. 17 a). This convolution has its maximum at the point with maximum overlap of the signals. As the point spread functions of the two signals are assumed to be the same this point represents the lateral distance between the two NV centres.

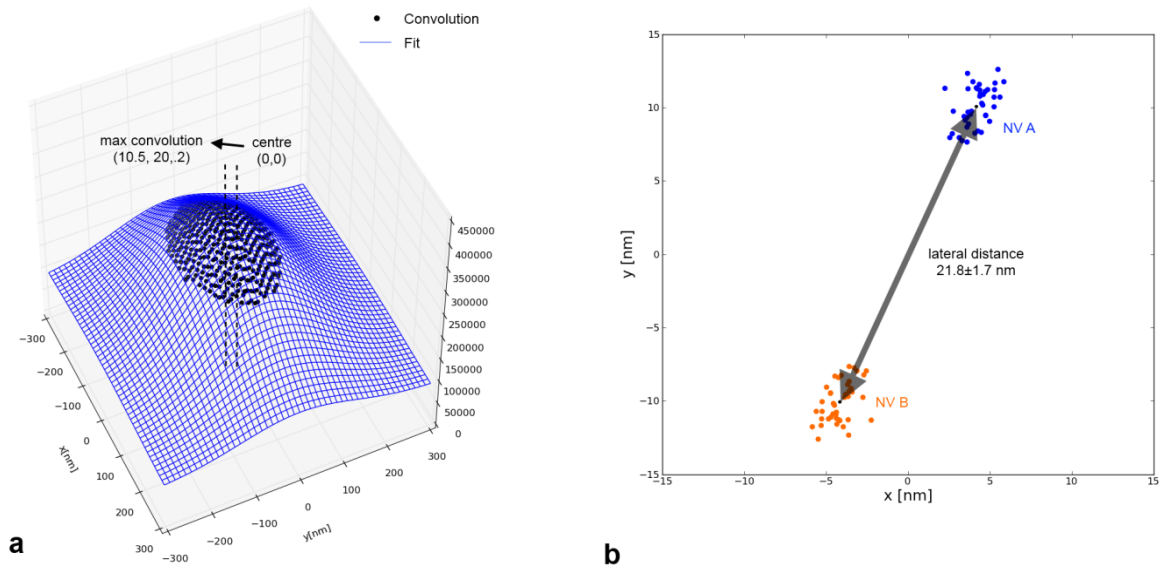

**Supplementary Figure 17 |** **a,** Convolution of the two difference signals. The elliptical Gaussian is fitted to the cap of the convolution in order have a better confidence in fit parameters for the position. The difference between the origin and the maximum of the convolution represents the distance of the NV centres. **b,** The lateral distance of the investigated pair is $\Delta r_{x/y} = 21.8 \pm 1.7$ nm. This is the mean value of 42 measurements. The error is the standard deviation of the measured distance vectors.

The distance in the surface plane of the pair investigated in this work could be obtained by this method. The average of several measurements is $\Delta r_{x/y} = 21.8 \pm 1.7$ nm (see Supplementary Fig. 17 b). The given error is the standard deviation of the measured distance vectors. A confocal scan images along the optical axis are less accurate. We achieve better results by estimating the z-component from the dipolar coupling strength and the lateral position. This way we determined the absolute distance to be in the range of 22 to 24 nm.